\documentstyle[aps,prd,eqsecnum]{revtex}
\input psfig.sty
\def\be{\begin{equation}}
\def\ee{\end{equation}}
\def\ba{\begin{eqnarray}}
\def\ea{\end{eqnarray}}

\def\gta{\mathrel{
\hbox to 0pt{\lower
3pt\hbox{$\mathchar"218$}\hss}\raise 2.0pt\hbox{$\mathchar"13E$}}}

\def\lta{
\mathrel{\hbox to
0pt{\lower 3pt\hbox{$\mathchar"218$}\hss}\raise 2.0pt\hbox{$\mathchar"13C$}}}

\begin{document}

\draft

\title{The fate of chaotic binaries}

\author{Janna Levin}
\address{DAMTP, Cambridge University,
Wilberforce Rd, Cambridge CB3 0WA}
\address{and}
\address{The Blackett Laboratory, Imperial College of
Science, Technology \& Medicine, South Kensington, London SW7 2BZ}
\address{e-mail: j.levin@damtp.cam.ac.uk}

\twocolumn[\hsize\textwidth\columnwidth\hsize\csname
           @twocolumnfalse\endcsname

\maketitle
\widetext

\begin{abstract}

A typical stellar mass black hole with a lighter companion is shown
to succumb to a chaotic precession of the orbital plane.  
As a result, the optimal candidates for the 
direct detection of gravitational waves by Earth based interferometers
can show irregular modulation
of the waveform during the last orbits before plunge.
The precession and the
subsequent modulation of the gravitational radiation
depends on the mass ratio,
eccentricity, and spins.  The smaller the mass of the companion,
the more prominent the effect of the precession.
The most important parameters
are the spin magnitudes and misalignments.  If the spins are small and 
nearly aligned with the orbital angular momentum, then there will be no chaotic
precession while
increasing both the spin magnitudes and misalignments
increases the erratic precession.  
A large eccentricity can be induced by large, misaligned spins but
does not seem to be required for chaos.
An irregular precession of the orbital plane will generate irregularities
in the gravitational wave frequency but may have a lesser effect on the 
total number of cycles observed.

\end{abstract}

\medskip
\noindent{04.30.Db,97.60.Lf,97.60.Jd,95.30.Sf,04.70.Bw,05.45}
\medskip
]

\narrowtext
 
\setcounter{section}{1}

Merging black hole binaries are potent sources
of gravitational waves
and are among the most promising targets for direct
detection by the future interferometric 
observatories.
Black hole mergers, if sufficiently abundant, are likely to be
the most common compact binary merger to be detected.
If the black holes are rapidly spinning, then the orbit
can be extremely irregular, even chaotic, bearing significant
implications for gravitational wave searches 
\cite{{maeda},{me1},{bhs},{bc},{carl},{cf},{me0}}.
An earlier {\it Letter} \cite{me1} 
identified chaos in relativistic, spinning binaries in
a somewhat abstract discussion.  In this article, the intention is
instead to provide a more concrete discussion with less emphasis
on formal chaos.
What is observationally important 
is visibly irregular motion.
Taking this attitude,
a specific astrophysical model is followed through successive
stages in order to gauge when 
irregular motion will occur within the LIGO/VIRGO bandwidth.
Specifically, we investigate the orbits of a maximally spinning
black hole with a lighter companion.  

Certain binary star systems are fated to evolve into black hole binaries.
The orbits of these long lived binaries
have sufficient time to circularize
before entering the LIGO/VIRGO bandwidth
as angular momentum is lost to gravitational waves.
An archive of circular templates is accruing for various
binary parameters.  Yet the merger rates of these evolved binaries
are predicted to be too low to ensure detection by the first two 
generations of LIGO detectors.  A more promising detection rate is 
predicted for dynamical binary black holes; that is, binaries formed 
by the dynamical capture of one black hole by another in dense stellar
systems \cite{pzm}.  The merger rate is expected to be about
$1.6\times 10^{-7}$/yr/Mpc${}^3$, which exceeds neutron star merger rates
as well. 
Dynamically formed black hole binaries should have
a distribution of eccentricities
and short orbital periods with masses in the range of
$\sim (5-15) M_\odot$ \cite{pzm}.
A binary
with masses $m_1=15M_\odot$ and $m_2=5M_\odot$ for instance 
will emit gravitational waves with a frequency 
within the optimal LIGO bandwidth of 
$f\sim (10-10^2)$ Hz for radial separations $r\lta 50 m$
where units of total mass $m=m_1+m_2$ are used.
These provide natural values
for the mass and radii ranges to investigate.  
The heavier black hole is taken to 
have maximal spin $S_1=m_1^2$ (spin period 
$P\sim 3\times 10^{-4} s $ for a $15M_\odot$ black hole).  
Unlike pulsars, 
black holes are 
expected to essentially maintain the spin they are born with 
\cite{mw} through
most of the
inspiral.  This canonical BH/BH pair can precess chaotically
any time the trajectory transits near the underlying homoclinic orbits
of Refs.\ \cite{loc}. Homoclinic orbits are purely relativistic,
a consequence of nonlinearity, and unstable. They have
the essential features for the onset of chaos \cite{{bc},{loc}}
when the bodies spin. Still, having said this, it is not clear
that chaos will be confined to this region of phase space.

A BH/NS binary with typical parameter values of
$m_1=10M_\odot$ and $m_2=1.4M_\odot$ follows trends similar to 
the BH/BH binaries.  
The explosive evolution of 
stellar progenitors which populate BH/NS pairs delivers large kicks to
the objects and leads to large spin misalignments \cite{kalogera}.
It is still unclear whether the population of such pairs is too sparse
to expect a detection.  Since it is only the mass ratio which enters
the equations, either of these cases can be scaled to represent the 
dynamics of much more massive systems which will be visible to
LISA \cite{reviews}.

We look for chaotic behaviour when energy is conserved and the
radiation
reaction is turned off.  
For an individual orbit, chaos manifests itself as the unpredictable
precession of the orbital plane. 
When considering all possible orbits, chaos manifests itself as an
extreme
sensitivity of the orbital precession to initial conditions so two 
neighboring orbits may live out very different precessional histories.
The implication is that there is a theoretical limit on how well we
can predict the orbit and therefore the waveform of the emitted gravitational
radiation
\cite{{maeda},{me1}}.
Dissipation from the emission of gravitational
waves is then included.  Irregular motion in the dissipative system
is understood in terms of the number of windings the pair executes in
a region of phase space which is chaotic for the underlying
conservative
system.

The regularity of the orbit will be effected by several parameters:
the mass ratio $\beta=m_2/m_1$, the magnitudes of the spins, the spin alignment
with respect to the orbital plane, the eccentricity of the orbit,
and the radius of the orbit at the time of detection.
As is already clear
from Ref.\ \cite{me1}, motion in the conservative system 
becomes more irregular
the larger the angle the spin makes with the perpendicular to the orbital
plane.
The importance 
of three other parameters is evaluated here:
(1) the binary mass ratio $\beta=m_2/m_1$, (2) the magnitude of the second
spin $\vec S_2$, and (3) the eccentricity of the orbit.

The conclusions in brief for the three parameters varied below are the
following:
(1)  The mass ratio primarily effects the cone of precession.
The smaller the mass
ratio
$\beta=m_2/m_1$, the larger the angle subtended by the orbital
plane and the larger the modulation of the gravitational waves
\cite{{acst},{kidder}}.
(2)  There can be chaotic motion when the massive black hole
spins rapidly even if the companion has no spin.  Still, the larger the 
magnitude of the second spin (and the misalignment), the more
irregular
the motion.
(3)  Eccentricity is a consequence of large, misaligned spins and
therefore it is difficult to separate cause and effect here. Still, it is 
clear that eccentricity alone is not responsible for chaos.

\section{Equations of motion and Spin Precession}

The Post-Newtonian (PN) expansion of the relativistic two body
problem leads to a system of equations describing the fate of spinning
binaries \cite{pn}.  
The PN expansion converges slowly to the fully relativistic
description \cite{kww}.  For this reason, it is a poor approximation
at small separations.  Despite its shortcomings, the PN expansion
does give the qualitative features of a relativistic system
such as nonlinearity, the existence of unstable circular orbits
\cite{kww},
homoclinic orbits \cite{loc}, and spin precession.  Since these
are the ingredients for chaotic dynamics, the qualitative behaviour
should persist in a more accurate approximation, although the quantitative 
conclusions are subject
to change (see for instance the improved technique of Ref.\ \cite{damour}). 

It is worth emphasizing that 
approximations can introduce
chaos when the exact system is truly regular.  One might worry that the
error
at 2PN order has introduced chaos which would be removed if we knew the
full 
equations of motion without approximation.  However, 
the relativistic two-body problem is
likely to be more irregular at higher orders as the nonlinearities
of general relativity are more accurately represented, not less.  
One might even 
be inclined to take the extreme resistance of the relativistic
two-body problem to solution
as evidence, or at least confirmation, of nonintegrability.

The validity of the PN expansion is not questioned further and 
the equations are treated as a self-contained
dynamical system.
In the PN scheme,
the orbit evolves according to the force equation
	\be
	\ddot{\bf \vec r }={\bf \vec a_{PN}} +{\bf \vec a_{SO}}+{\bf \vec
	a_{SS}}+{\bf \vec a_{RR}} \label{eom1}
	\ee
in center
of mass harmonic coordinates \cite{kidder}.
The 
acceleration is due to Post-Newtonian (PN) effects, spin-orbit (SO)
and
spin-spin (SS) coupling and 
radiative reaction (RR).
The explicit 
form of ${\bf \vec a}$ can be found in
the appendix.
The spins also precess due to the relativistic frame dragging 
and Lens-Thirring effect.  The precession equations are
	\be
	\dot{\bf \vec S_1} = {\bf \vec \Omega_1}\times {\bf  \vec S_1 }
	\quad  , \quad
	\dot{\bf \vec S_2 }={\bf  \vec \Omega_2}\times {\bf  \vec S_2} 
	\label{prec1}\ee
with
	\begin{eqnarray}
	{{\bf \vec \Omega_1}} = {1 \over r^3} \biggl [ &&
	\left (2+{3 \over 2}
	{m_2 \over m_1}\right) {\bf \vec  L_N} 
	-{\bf \vec  S_2 } 	
	+ 3({ {\bf \hat n} \cdot {\bf \vec S_2}})
	{ {\bf \hat n} } \biggr ]
	,
	\end{eqnarray}
and
	\begin{eqnarray}
	{{\bf \vec \Omega_2}} = {1 \over r^3} \biggl [ &&
	\left (2+{3 \over 2}
	{m_1 \over m_2}\right) {\bf \vec  L_N }
	-{\bf \vec  S_1}  	
	+ 3({ {\bf \hat n} \cdot {\bf \vec S_1}})
	{ {\bf \hat n} } \biggr ].
	\end{eqnarray}
The spins precess with constant magnitude although the total
spin ${\bf \vec S}={\bf {\vec S}_1}+{\bf {\vec S}_2}$ may not have constant
magnitude.

\begin{figure}
\centerline{\psfig{file=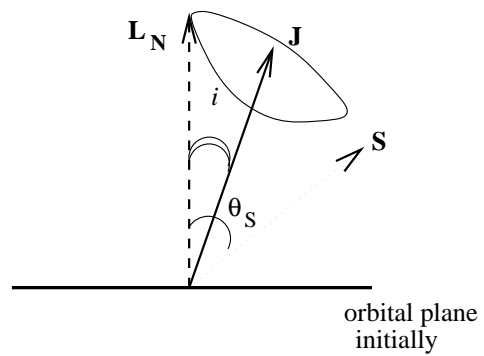,width=2.5in,angle=-90}}
\vskip 25truept
\caption{
A schematic drawing of the inclination angle $i={\rm arccos}
\left (
{\bf \hat L_N}\cdot {\bf \hat J}\right )$ and the angle 
$\theta_S={\rm arccos}\left ({\bf \hat L_N}\cdot{\bf \hat S}\right )$.
The orbital plane traces out a band as in figs. \ref{m.14r20} and
\ref{m.3r20} as the Newtonian angular momentum precesses about ${\bf \vec J}$. 
\label{angles}}  \end{figure} 

The
orientation of the orbital plane is defined by 
the Newtonian orbital angular momentum
	\be
	{{\bf \vec  L_N}} \equiv \mu ({{\bf \vec  r} \times {\bf \vec v}})
	\label{lnewt}
	\ee
with the reduced mass $\mu=m_1m_2/m$ and the total mass $m=m_1+m_2$.
Spin precession, generates a precession of the orbital plane
(fig.\ \ref{angles}).  This 
can most easily be seen by noting that 
to 2PN-order the
total angular momentum ${\bf \vec J}$ is conserved
with ${\bf \vec J}={\bf \vec L}+{\bf \vec S}$.
The
orbital angular momentum  ${\bf \vec L}$
can be split into two pieces,
${\bf \vec L}=\alpha {\bf \vec L_N}+{\bf \vec L_{SO}}$ as in 
eqns.\ (\ref{split}) and (\ref{split3}).  The term ${\bf \vec L_{SO}}$ 
is due to spin-orbit coupling
and $\alpha $ contains Post-Newtonian corrections.
To 2PN order $\dot {{\bf \vec J}}=0$ and so
	\be
	\dot {\bf \vec  L}\sim -\dot {\bf \vec S}.
	\ee
The magnitude of the orbital angular momentum is {\it not}
constant.  
Further, the precession of the orbital plane can be much more
complicated than the precession of ${\bf \vec S}$:
$\left (\alpha {\bf \vec L_N}\right )^{\dot {}}\sim -\dot {\bf \vec S}-
\dot {\bf \vec L_{SO}}$.
The orbital plane therefore does not just carve out a simple cone as it 
precesses around the direction of ${\bf \vec J}$.  Instead the plane 
tilts back and forth as it precesses.
The motion becomes chaotic as the tilting back and forth becomes
highly irregular in the Hamiltonian system.
When radiation reaction is included, the degree of irregularity in the
precession of the orbital plane can be estimated in terms of how 
many windings the pair spends near the chaotic region of the underlying
conservative system. 

\begin{figure}
\centerline{\psfig{file=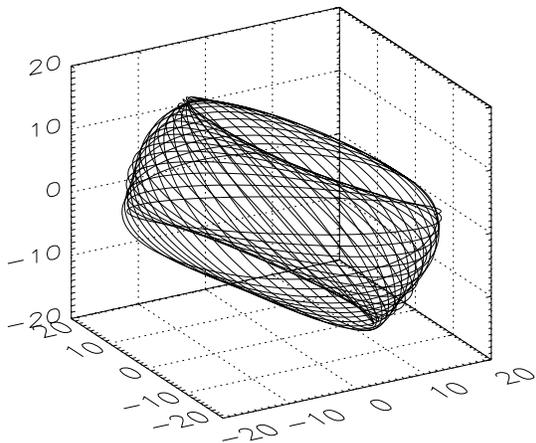,width=4.in}}
\caption{A  three-dimensional view of a regular 
orbit with $\beta=m_2/m_1=1.4/10$, $S_1=m_1^2$ and $S_2=0$.
The initial angle $\theta_1={\rm arccos}({\bf \hat L_N}\cdot{\bf \hat S_1})
= 45{}^o$.  The initial conditions for the orbit are $r/m=20$ and
$r\dot \phi=0.209$.
\label{m.14r20}}  \end{figure}

Fig.\ \ref{m.14r20} shows an example of the trajectory for the
center of mass of a BH/NS pair.  The motion is plotted in 
three-dimensions to illustrate the precession of the orbital plane.

The precession, whether regular or irregular, modulates the emitted
gravitational waves.
The response of a detector on Earth to an impinging gravitational wave
can be parametrized as
	\be
	h=F_+h_++F_\times h_\times
	\ee
where $h_+$ and $h_\times$ are the two gravitational wave
polarization states and 
$F_+$ and $F_\times$ are the antenna patterns.  A radiation coordinate
system can be defined such that the polarization axes are fixed even
in the presence of precession \cite{chern}.  In such a coordinate system,
$F_+$ and $F_\times$ are constant.  However, the polarization states
$h_+$ and $h_\times$ depend on the inclination of the orbit and the
precessional frequency in a complicated way \cite{kidder}.
For a circular orbit, the detector response can be written
	\be
	h=A\cos(2\Phi-\delta)\label{simp},
	\ee 
where the higher harmonics have been ignored for simplicity.
The amplitude $A$ and polarization phase $\delta$
depend on the binary's location, orientation, and precession
\cite{{acst},{kidder}}.
For an elliptical orbit, $h$ has terms of the form
$\cos(\Phi),\cos(2\Phi)$, and $\cos(3\Phi),$ at quadrupole order so 
the gravitational wave spectrum shows oscillations at once,
twice and three times the orbital frequency.

Precession of the orbital plane will (1) modulate the amplitude in eqn.\
(\ref{simp}) (2) modulate the polarization phase 
$\delta $ and therefore the frequency of the gravitational waves
and (3) contribute to the
overall accumulated phase by changing the inspiral lifetime.  Any extreme
sensitivity to initial conditions will most likely have the largest effect
on the modulation of the amplitude and frequency of the gravitational waves.
The overall accumulated contribution to the number of cycles in the observed
waveform will certainly be effected by the general bulk precession but
may be less sensitive to the irregularity of the precession until
the very final stages of coalescence.
The reason for this is that at the radii accessible to the interferometers,
the irregularity seems to predominately 
effect the orientation of the orbit with a lesser effect on the net
orbital velocity
$\vec \omega = \dot \theta \hat \theta +\sin\theta \dot \phi \hat \phi$.
LIGO/VIRGO aim to observe gravitational waves by accurately measuring the
accumulated phase defined as
	\be
	\Psi=\int \omega dt=\int {\omega \over \dot \omega}d\omega
	,
	\ee
which may only have a small correction from the irregularity
of the precession.  Irregular motion will effect the phasing, that is 
the gravitational wave frequency, and the amplitude of the wave.  
The number of cycles in the accumulated
phase may therefore be determined by the average bulk
behaviour
of the precession.
Although this conjecture requires further scrutiny, we can use the 
approximations  of Refs.\ \cite{{acst},{kidder}} to provide a rough figure
for the number of wave cycles.  In Ref.\ \cite{acst}, circular
Newtonian orbits
were studied.  The effect of spin precession was isolated without including
spin-orbit acceleration terms in the equations of motion
in Ref. \cite{acst}.  This separation
of the spin precession equations and the equations of motion removed
any possibility of chaotic coupling.
However,  the gross features can be fairly represented by this approach.
They estimate that the change in the total number of cycles amounts to 
about twice the total number of precessions.  For the binary black holes
of size $15 M_\odot +5M_\odot$, there can
be $\sim 10 - 15$ precessions in the observable band
and so there can be $\sim 20-30$ 
additional cycles (depending on spin orientation, magnitude and eccentricity
of the orbit).
In Ref.\ \cite{kidder}, the contribution to the total number of orbits
over the entire LIGO bandwidth
was also estimated to be $\sim {\cal O}(20)$ for black hole binaries
with total mass $\sim 20 M_\odot$.  The number of additional cycles could
quadruple for NS/NS masses \cite{kidder}

It has been argued that matched filtering will be a poor means by
which to observe BH/BH orbits in the chaotic regime
\cite{{me1},{scott}} and that other cruder methods will be employed.
We contribute to this debate only by indicating when
irregularity will influence detectability.  It has also been emphasized
in 
Ref.\ \cite{scott}, that the PN expansion is not sufficient to
accurately model templates for $r \lta 15 m$.  Nonetheless, the higher
order
contributions will incorporate stronger relativistic effects and
therefore
more nonlinearity which should only lead to more irregular motion.
Since we are trying to provide a physical picture of the general
trends and dependences, we continue to use the PN expansion.

\section{chaos}
\label{chaos}

Chaos in relativistic systems is notoriously difficult to quantify
\cite{{njt},{mixm}}.  While formal definitions of chaos are of little
interest to data analysts when irregularity regardless poses
a hindrance, 
the tools of chaos are nonetheless
critical to survey the system for
endemic irregularity.
In Ref.\ \cite{me1}, chaos in the spinning black
hole problem was identified through the method of fractal basin
boundaries
\cite{{carl},{cf},{me1}}.  In this section irregularity of
an individual orbit is discussed and the power of the fractal basin
boundary technique is utilized.
It is worth emphasizing that the fractal basin boundary method
allows one to scan huge numbers of orbits and therefore
provides an invaluable tool to survey phase space for chaos.

The equations of motion and spin precession equations are treated as 
any other dynamical system.  
In the first instance
radiation reaction is omitted and chaos is handled
in the conservative system.
Dissipation is treated in \S \ref{diss}.
The 2PN equations of motion can be derived from a Lagrangian
\cite{{dd},{kww2}} and therefore can be derived from a Hamiltonian.
The
coordinates ${\bf \vec S_1}(t)$ and ${\bf \vec S_2}(t)$ can be treated
as external time-dependent perturbations to the integrable system
with the equations of motion supplemented by the precession eqns.\ 
(\ref{prec1}).
The system can therefore be treated as a time-dependent
Hamiltonian system $H({\bf \vec r},{\bf  \vec v},{\bf 
\vec S_1}(t),{\bf \vec S_2}(t))$.  

Chaos is well defined for Hamiltonian systems. Chaos is synonymous 
with nonintegrability.  
Regularity is synonymous with integrability.  
In a
Hamiltonian system with
$N$ degrees of freedom
$q$ and $N$ conjugate momenta $p$,
integrability prevails when there are $N$
independent constants of the motion.
The constants of motion must
also be in involution; that is, the Poisson bracket of any constant
with the others
vanishes: $[C_i,C_j]=0$.  The $2N$-dimensional phase space can then be
reduced to motion on an $N$-dimensional torus.  This is most easily 
represented with a canonical change of coordinates to action angle variables
$(I,\Theta)$ such that each of the $N$ momenta $I$ 
is set equal to one of the $N$
constants of motion $C$.  The motion can then be made periodic in the
coordinate variable $\Theta$.  For $N=1$, degree of freedom, the
motion lies on a circle.  For $N=2$, the motion lies on a torus
and for arbitrary $N$, the motion lies on an $N$-dimensional torus
\cite{{ll},{ott}}.
In any other set of canonical variables
besides action angle variables, the mark of integrability is that 
the motion is confined to a smooth closed curve in $(p,q)$ and does not 
diffuse off that line.  If the
motion in $(r,\dot r)$ has diffused off of a smooth line
it is not restricted to a torus and 
the motion must be nonintegrable.

\begin{figure}
\centerline{\psfig{file=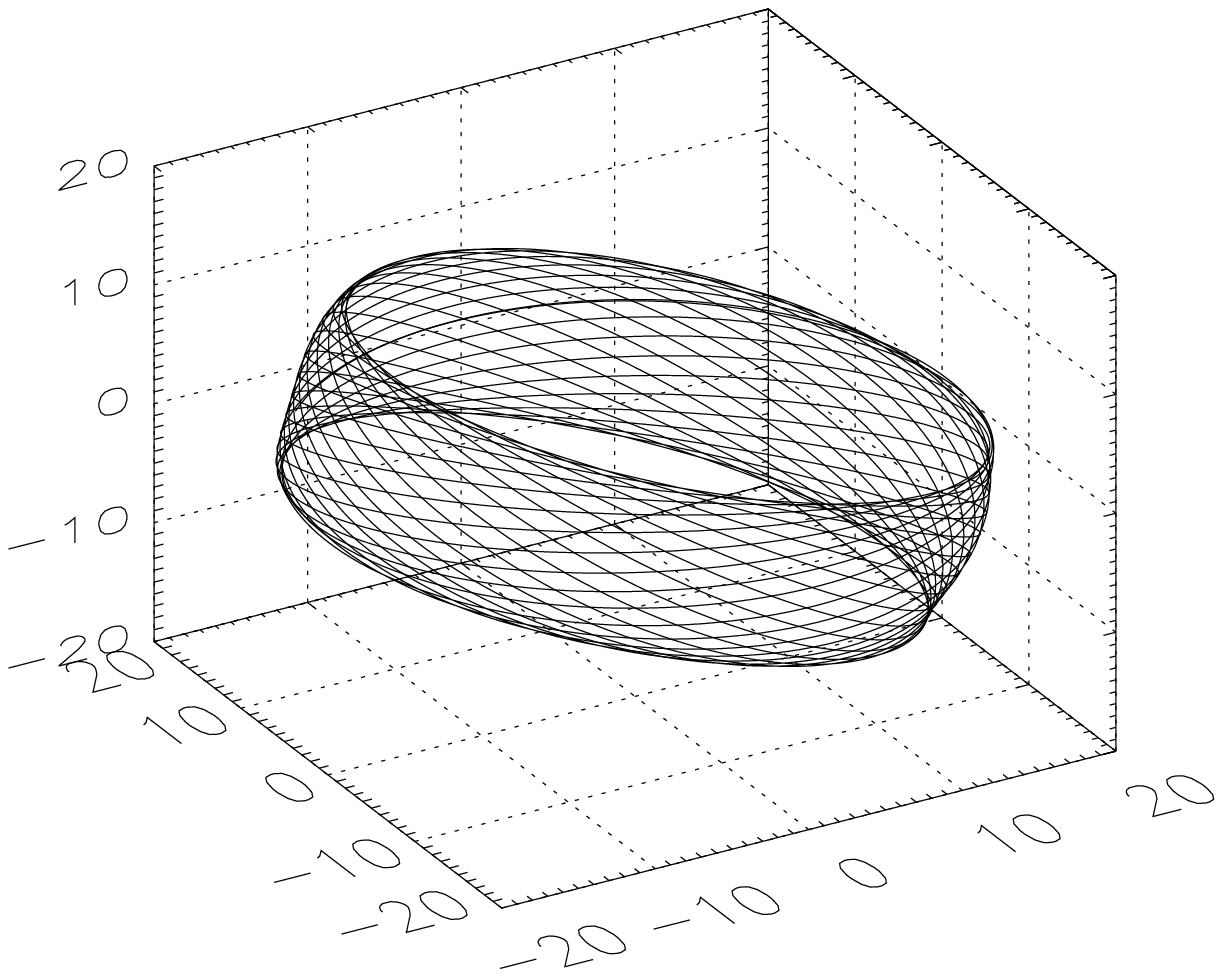,width=4.in}}
\centerline{\psfig{file=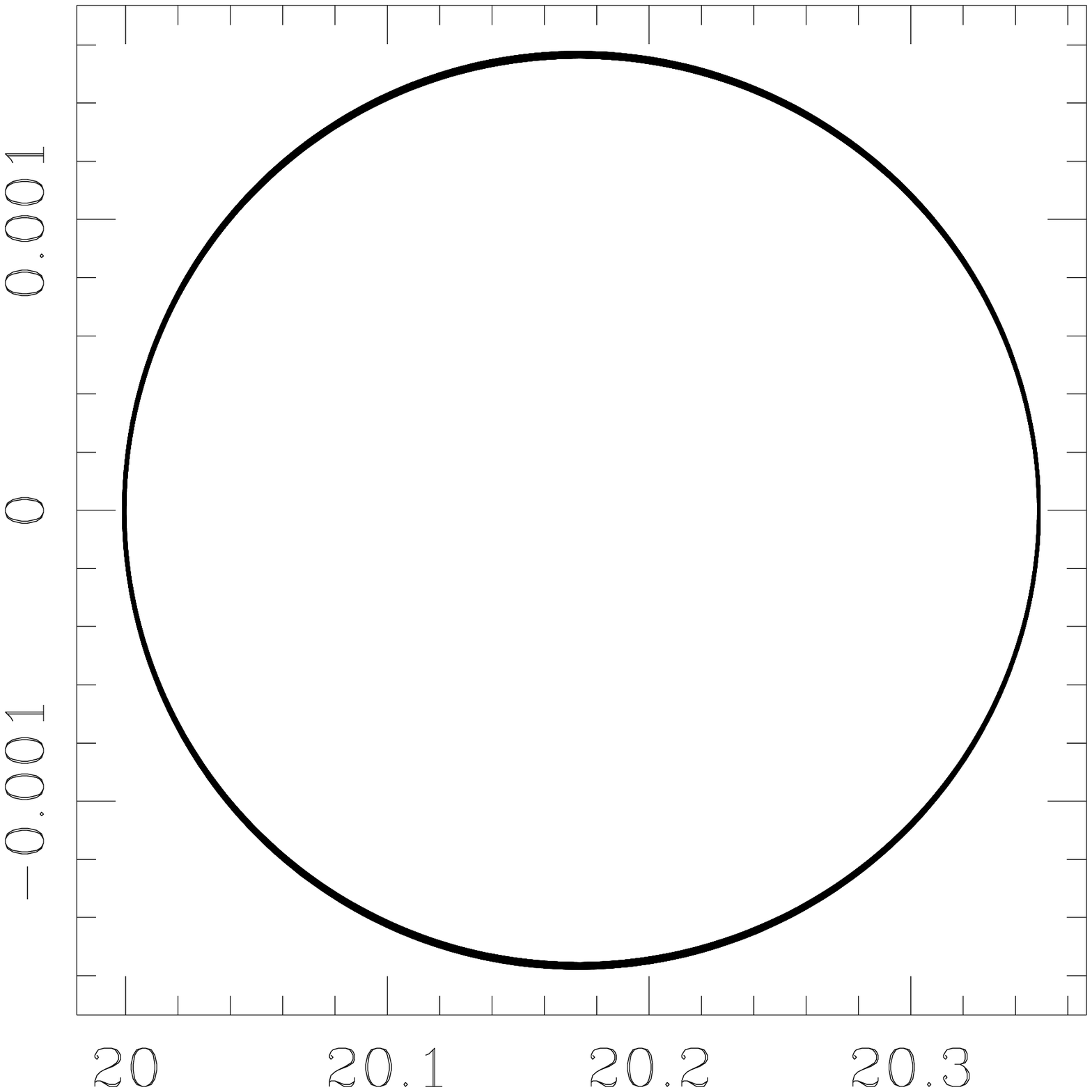,width=2.1in}}
\caption{A regular orbit with $\beta=m_2/m_1=1/3$, $S_1=m_1^2$ and $S_2=0$.
The initial angle $\theta_1={\rm arccos}({\bf \hat L_N}\cdot{\bf \hat S_1})
= 45{}^o$.  The initial conditions for the orbit are $r/m=20$ and
$r\dot \phi=0.209$.
Top:  three-dimensional view of the orbit.
Middle:  a projection of the orbit onto the $(r,\dot r)$ plane.
\label{m.3r20}}  \end{figure}

For coalescing binaries,
there are $N=3$ degrees of freedom, $(r,\theta,\phi)$.
When there are no spins, the phase space is $2N=6$-dimensional.  The
energy and angular momenta provide enough constants of the 
motion to restrict the trajectories to tori and there is no chaos
to 2PN order \cite{{me1},{loc}}.  Beyond 2PN order
on the other hand, the two-body problem is likely to be
chaotic even in the absence of spin.

When the bodies spin,
the dynamics can be reduced to a time-independent Hamiltonian system
by taking a Poincar\'e surface of section.  
This method of identifying chaos is less than ideal here. For the record, 
we will
take the time to explain the method and its shortcomings.
The Poincar\'e surface of section involves
plotting a point in $(r,\dot r)$
each time the orbit crosses a surface on which all of the other coordinates
are fixed.
If the collection of points defines a smooth curve, then the motion
is confined to a torus and is integrable.  If the points speckle the plane,
then the motion has diffused off of a torus and is nonintegrable.
The Poincar\'e reduction of phase space is most easily
demonstrated for the case of only one spinning body so that
${\bf \vec S}={\bf \vec S_1}$.  
Including spin, the phase space is $2N+3=9$ dimensional.
The
$4$ constants of motion, $E,{\bf \vec J}$,
restricts motion to a $5$-dimensional surface.
If we treat the Hamiltonian as periodic in the three coordinates
${\bf \vec S}(t)$, we can 
plot a point in 
$(r,\dot r)$ each 
time the orbit crosses
${\bf \vec S}(t)={\bf \vec S}(0)$ so that there are only two 
free coordinates remaining.  
In this way we can look for the destruction of tori and test
for nonintegrability.

A regular BH/BH binary 
with 
mass ratio $\beta=m_2/m_1=1/3$
is shown in fig.\ \ref{m.3r20}.
The more massive black hole has spin
$S_1=m_1^2$ with ${\rm arccos}({\bf \hat L_N}\cdot{\bf \hat S_1})=45^\circ$
initially.  The second spin is set to zero for this figure.
The top panel in fig.\ \ref{m.3r20} shows a three-dimensional view
of the orbit.  The middle panel shows a projection 
of the orbit in the $(r,\dot r)$ plane. This orbit is very nearly circular.
Note that if the orbit is exactly circular with only 
one spin then there can be no chaos since the dimensionality of the phase
space reduces to two \cite{me1}, which is not enough to support chaos. 
This orbit is regular as indicated by the absence of spreading
in the projection onto the $(r,\dot r)$ plane.  

\begin{figure}
\centerline{\psfig{file=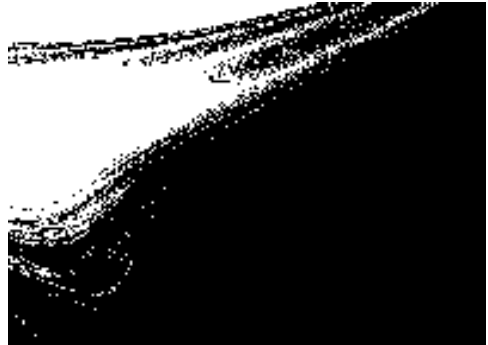,width=2.5in}}
\caption{
A fractal basin boundary scan in $(\dot r,r \dot \phi)$ varying over 
initial values in $0 <\dot r< 0.035$
and $0.425 < r\dot \phi<0.443125$.
All 40,000 orbits have a maximally spinning black hole and a lighter
companion with no spin ($\beta=m_2/m_1=1/3$, $S_1=m_1^2$ and $S_2=0$).
The initial angle is $\theta_1={\rm arccos}({\bf \hat L_N}\cdot{\bf \hat S_1})
= 95{}^o$ and initial separation $r/m=5$.
$200 \times 200$ orbits shown.
\label{fbbv}}  \end{figure}

\begin{figure}
\centerline{\psfig{file=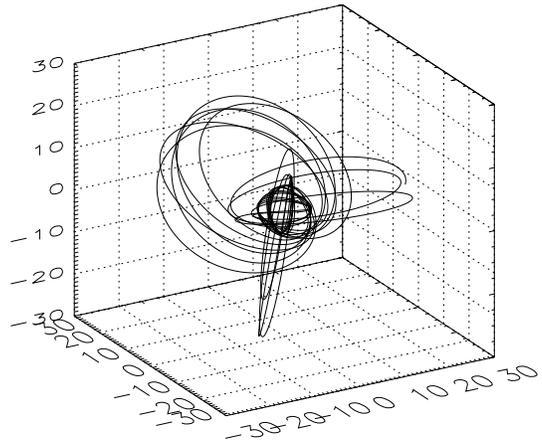,width=4.in}}
\centerline{\psfig{file=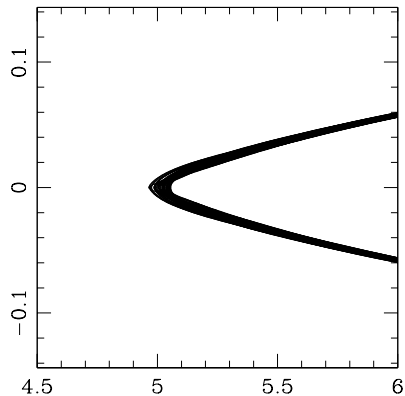,width=2.25in}}
\caption{
Top: A three-dimensional plot of an irregular orbit near the basin boundary of
fig.\ \ref{fbbv}
with initial velocities $\dot r =0.00105$ and $r\dot \phi=0.43075$.
Bottom:  A projection of the orbit onto the
$(r,\dot r)$.
\label{fbbvorb}}  \end{figure}

By contrast, chaotic orbits are shown in figs.\ \ref{fbbv} and
\ref{fbbvorb}.
Chaos is first identified by the method of fractal basin boundaries.
To build the basin boundary, 40,000 black hole binaries are evolved
varying only the initial velocities ($\dot r, r\dot \phi$). 
The initial condition is then color coded according to outcome:
black if the pair merge and white if the pair execute at least 50
or more orbits. (Increasing the required number of orbits will shrink the
white basins but will not eliminate structure at the boundaries.)
A fractal boundary signifies extreme sensitivity to initial conditions
and a mixing of orbits -- i.e., chaos.
The power of the fractal basin boundary method as a survey of large
collections of orbits is clear.  Orbits near the boundary will
be chaotic.

An irregular orbit selected near the fractal boundary is shown 
the top panel of fig.\ \ref{fbbvorb}.
The bottom panel of fig.\ \ref{fbbvorb} is a detail of the projected motion 
in the $(r,\dot r)$ plane which shows some threading of the orbit.
A full Poincar\'e section taken as described above confirms that
this is not an illusion from the projection. 
However, the numerical burden is extreme and impractical in nearly all 
cases of interest.
For this reason, we continue to rely on the fractal basin boundaries and
only take the projection onto the $(r,\dot r)$ plane as a crude guide
and not as proof of chaos.

In theory, the surface of section technique
can also be implemented in the case of two spinning bodies
since both are periodic under the precession.
In practice this can be difficult since one might have to wait
a very long time before ${\bf \vec S_1}(t)={\bf \vec S_1}(0)$ 
at the same time 
as ${\bf \vec S_2}(t)={\bf \vec S_2}(0)$.  
Alternatively, we can cheat 
and simply look at the 
projection of $(r,\dot r)$ in the
full phase space without taking a Poincar\'e section.  
{\it This is only used as a crude survey  tool.}
If the
projected motion lies on a simple closed curve, the motion is decidedly 
integrable.  If the motion lies on a threaded orbit (such as that of 
fig.\ \ref{fbbvorb}), then this indicates the motion {\it might} 
have diffused off of
a torus. \footnote{A fruitful analytic approach may be 
to treat the 
motion as an instance of Arnol'd diffusion \cite{ll}.
Even if $\dot r$ is small and ${\bf \vec S}$ changes slowly, their
presence
induces a coupling between $\dot\theta $ and $\dot\phi $ which
can 
lead to chaotic resonances and hence stochastic behaviour.
The future direction of this work is to interpret the chaotic
motion in terms of this slow
modulation diffusion \cite{ll}.
}

Another major shortcoming of the Poincar\'e surface of section method in this 
setting is that the 2PN constants of motion are only approximately conserved.
Therefore even if one is cautious, the spreading may be a result of the
approximation and not a true mark of the destruction of tori.
Since this projection is ambiguous, it is only used as a rough guide. 
For a firm
identification of chaos we rely on the unambiguous fractal basin boundaries.

\begin{figure}
\centerline{\psfig{file=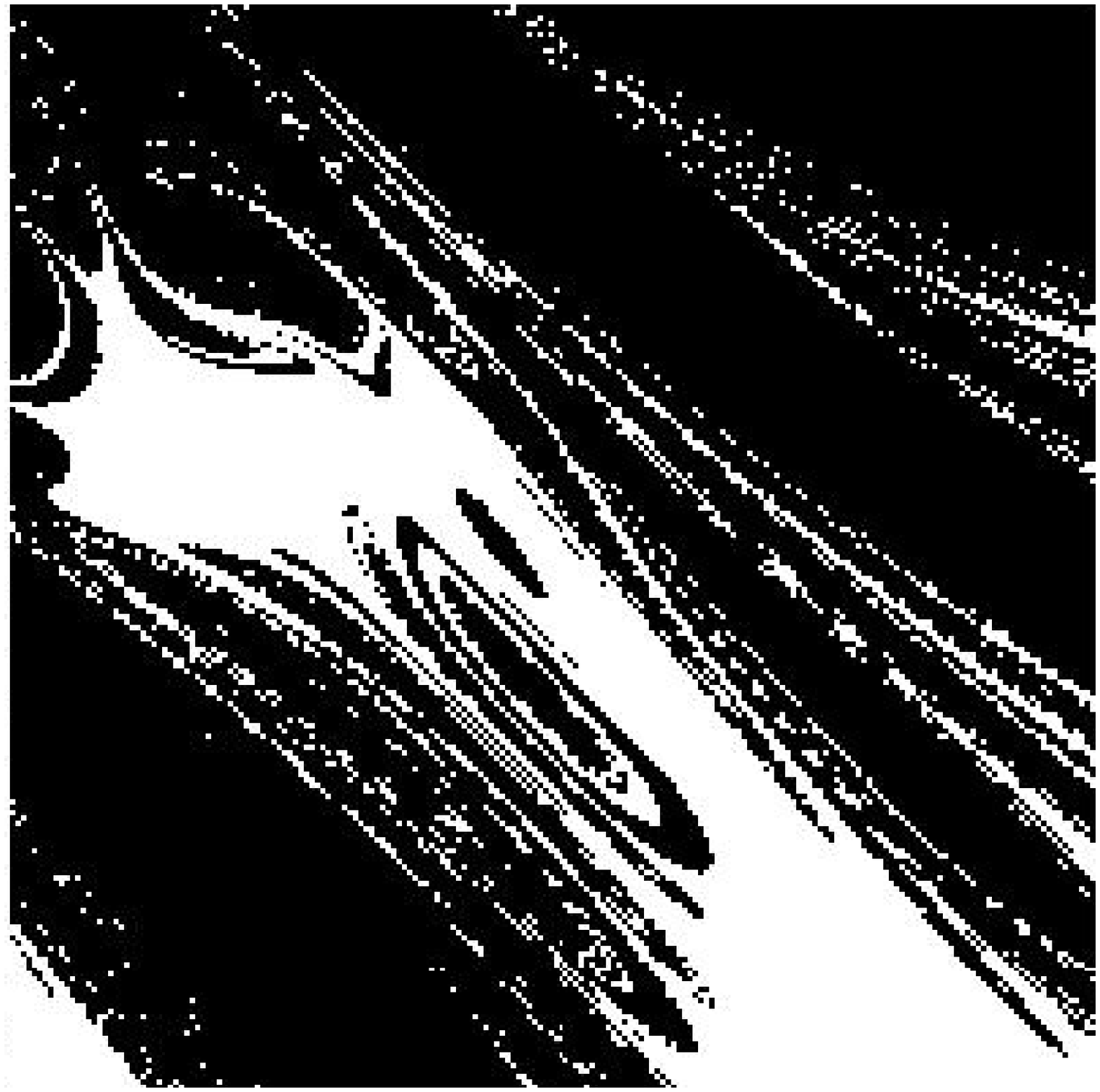,width=2.75in}}
\centerline{\psfig{file=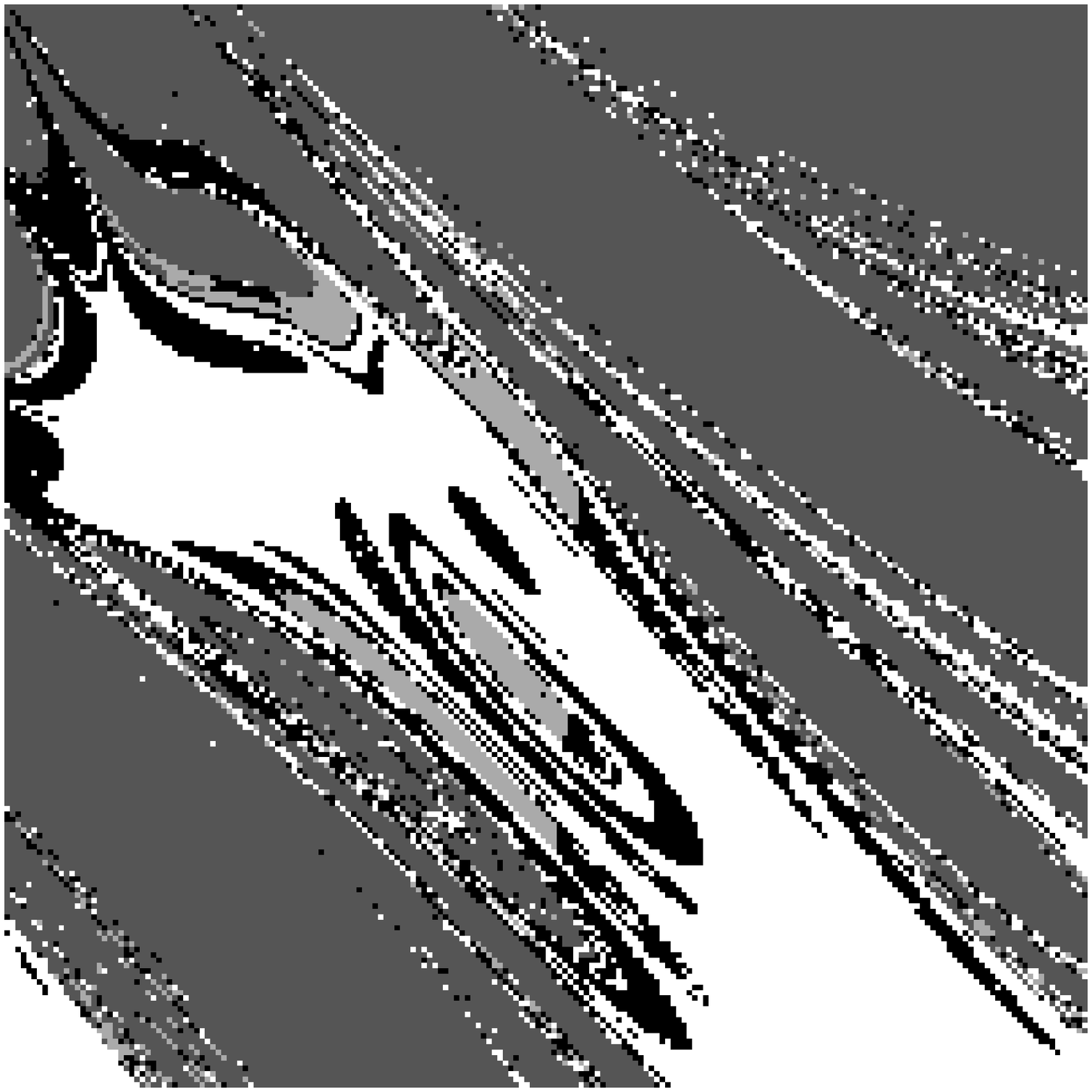,width=2.75in}}
\caption{All orbits have $m_2/m_1=1/3$, $r=5m$, and $r\dot \phi=0.45$.
The initial angles are varied in the $(\theta_1,\theta_2)$ plane over
the ranges
$140.375{}^o \le \theta_1 \le 157.5635{}^o$ and 
$45.8367{}^o\le \theta_2 \le 63.025411{}^o$.
Top: Figure from 
The fractal basin boundaries built up using the stable/merger
criteria.
Bottom:  The fractal basin boundaries built up using a critical winding
criterion. 
A resolution of $200 \times 200$ orbits.
\label{fbbn}}  \end{figure}

There are other outcomes that could be used to color code a plane in 
phase space and study the basin boundaries. A set of outcomes based on 
the number of windings a pair execute would be relevant to the 
search for gravitational waves.  In fig.\ \ref{fbbn}, basin boundaries for the same set of initial conditions are compared.
The binaries all have the
same initial condition except for the initial angular displacement of the
spins.
The top panel was originally
published in \cite{me1}. For that top panel, stable/merger outcomes
are used.  The initial condition in the $(\theta_1,\theta_2)$ plane is 
color coded white if the pair executes at least 50 orbits and black
if the pair merge in under 50 orbits. The lower panel uses a winding 
criterion:
the initial condition in the $(\theta_1,\theta_2)$ plane is color coded
white if the pair execute more than 50 orbits before merger, 
black if they merge after executing more than 36 orbits (but less than 50), 
dark grey if the pair merge 
having executed 36 orbits, and light grey if the pair merge having executed
less than 36 orbits.
The mixed basin boundaries show that there is some unpredictability
in the number of orbits executed in the conservative system.

We discuss dissipation in \S \ref{diss}. We point out
here that when dissipation is
included, there is less fractal structure when the critical windings 
are used as outcomes.  This indicates that with dissipation included the number
of windings executed by a binary might be predictable. However there is still
structure in the basin boundaries when other outcomes are used indicating
that not all features of the wave form will evolve predictably.

\section{The binary mass ratio}

The binary mass ratio $\beta=m_2/m_1$ primarily effects the cone
of precession. 
The lighter the relative mass of the
companion, the larger the band in which the orbital
plane will precess and the larger the corresponding modulation of the
waveform \cite{{acst},{kidder}}.
The orbital plane will precess around ${\bf \vec J}$ with an angle
of inclination
	\be
	\cos{i}= {\bf \hat L_N}\cdot {\bf \hat J}
	={{1+(S/L_N)\cos\theta_S}\over
	\sqrt{1+(S/L_N)^2+2(S/L_N)\cos\theta_S}}
	\label{incline}
	\ee
where $\cos\theta_S\equiv{\bf \hat L_N}\cdot{\bf \hat S}$ 
as in fig.\ \ref{angles}
and 
the total angular momentum has been used to lowest order, ${\bf \hat J}\sim 
{\bf \hat L_N}+{\bf \hat S}$.
The angle subtended changes as the angle between ${\bf \vec L_N}$ and
${\bf \vec S}$ changes.  This leads to the additional tilting back and 
forth on top of the simple precession.
The ratio $S/L_N$ can be estimated by taking $L_N \sim \mu (mr)^{1/2}$
for a nearly circular orbit.  
Consider the extremes when the more massive object spins with 
$S_1=m_1^2$ to get an upper range and the opposite regime when 
the lighter star spins with $S_2=m_2^2$ to get a lower range:
	\be
	\left (m_2\over m_1\right )
	\left (m/r\right )^{1/2} \le
	{S\over L_N} \le \left ({m_1\over m_2}\right ) 
	\left (m/r\right )^{1/2} \ \ .
	\label{lims}
	\ee
For very small $S/L_N$, the precession cone angle is so tight
that modulations of the gravitational wave signal will not be substantial.
Notice from the left-hand side of eqn.\ (\ref{lims})
that if only the lighter object spins then
$S/L_N$ is small for all $r$.  This may explain why Ref.\ \cite{maeda} found 
chaos for a test particle around a Schwarzschild black hole only
if the light companion had a spin $S_2 > 0.64 m_2 m_1$ for $m_1 \gg m_2$,
which exceeds the maximal spin of $S_2=m_2^2$.
The 2PN expansion to the two-body problem allows both black holes to spin
and precess.  When the more massive object spins maximally, then 
$S/L_N$ is large for $m_1 \gg m_2$.  For a $10:1$ mass ratio, $S/L_N\sim 1$ for
$r \sim 100 m$.  As a result, for large mass ratios, the orbital plane
subtends a larger angle at a given radius and modulation will be 
correspondingly larger \cite{{acst},{kidder}}.  
Consider fig.\ \ref{m.14r20} versus fig.\ \ref{m.3r20}.
In both figures, the more massive BH spins maximally ($S_1=m_1^2$)
with a spin displacement of
${\bf \hat L_N}\cdot {\bf \hat S}=\cos(45^\circ)=1/\sqrt{2}$. 
In both figures, 
the companion has no spin.  The difference between the two figures
is the mass ratio.
In fig.\ \ref{m.14r20}, the mass ratio is $\beta=1.4/10$ and
the angle subtended at $r=20$ is $i\sim 28^\circ$.
In fig.\ \ref{m.3r20}, the mass ratio is $\beta=1/3$ and the angle
subtended at $r=20$ is 
$i\sim 18^\circ $.  The band subtended by the precessing plane is 
larger for the smaller mass ratio, although it is substantial in 
both cases.

We can invert eqn.\ (\ref{incline}) using the right-hand-side of eqn.\
(\ref{lims}) to write the radius at a given inclination,
 	\ba
	&{}&\left ({r \over m}\right )^{1/2} \sim
	 \left (m_1 \over m_2\right )\times\nonumber\\
	&{}&\left [{\cos^2i-\cos^2\theta_1
	\over \cos\theta_1(1-\cos^2i)+ \sqrt{\cos^2i
	(1-\cos^2i)(1-\cos^2\theta_1)}}\right ]
	\label{crud}\nonumber
	\ea
with $\cos \theta_1\equiv {\bf \hat L_N}\cdot {\bf \hat S_1}$.
We could take this as an indicator for the radius at which 
precession becomes important.
Letting $i\sim 15^\circ $ and $\theta_1 \sim 45^\circ$, then
	\be {r\over m} \sim 4 \times \left (m_1 \over m_2 \right )^2
	.
	\ee
For circular motion, $L\sim \mu(mr)^{1/2}$ and $m\dot \phi\sim
\left (m/r\right )^{3/2}$ and the frequency of the emitted 
wave, $f\sim \dot \phi/\pi$ is
roughly
	\be
	f_{}\sim 9 \times 
	10^3{\beta^3}\left ( M_\odot\over m\right )\
	{\rm Hz} .
	\label{feq}
	\ee
For a BH/BH binary with $\beta \sim 1/3$ and $m=20 M_\odot$, then 
$f\sim {\cal O}(10)$ Hz when the spin precession angle
opens to $i\sim 15^\circ$ and the effects on the gravitational wave
should be noticeable.
For a BH/NS binary with $\beta \sim 1.4/10$ and $m=11.4$, then
$f\sim {\cal O}({1})$ Hz when the spin precession angle
opens to $i\sim 15^\circ$.  To emphasize, precession will be important
for these and larger frequencies as the pairs sweep through the 
interferometer's
bandwidth. 

\begin{figure}
\centerline{\psfig{file=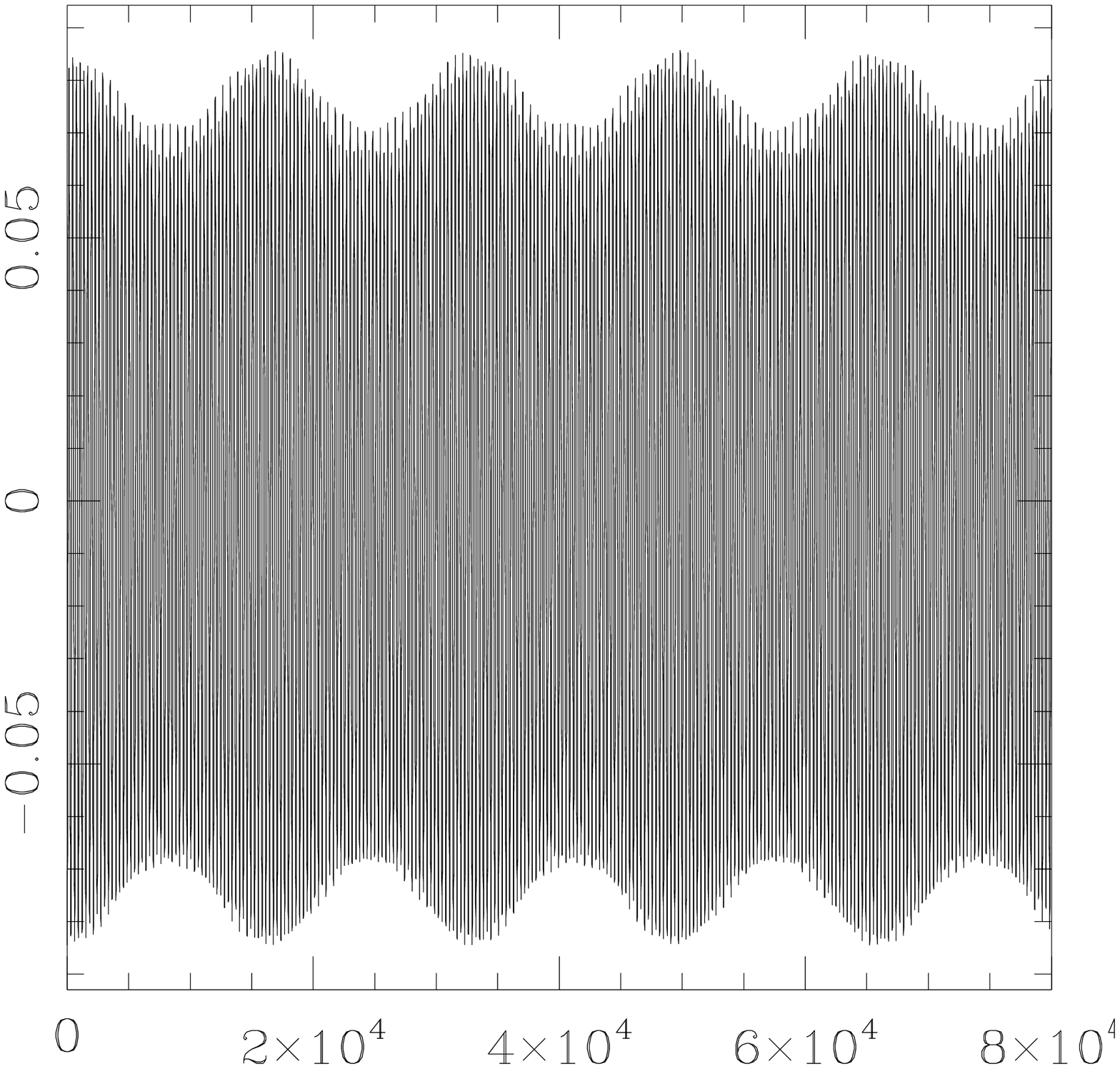,width=2.1in}}
\centerline{\psfig{file=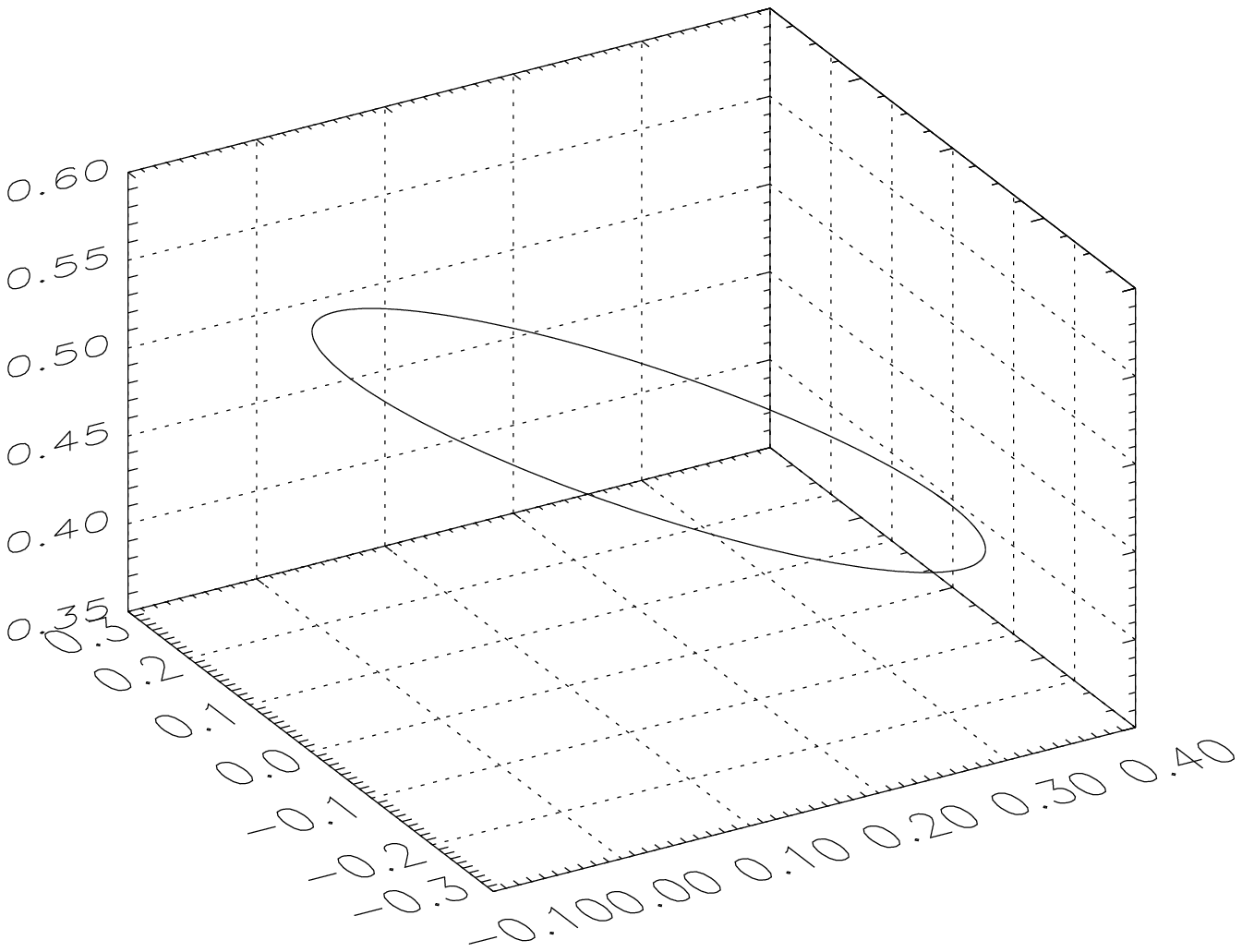,width=3.2in}}
\caption{The waveform, Newtonian angular momentum and spin for the 
orbit of fig.\ \ref{m.3r20}.
Top:  
The waveform $h_+$.
Bottom:  three-dimensional view of the precession of ${\bf \vec S_1}$.
\label{spinm.3r20}}  \end{figure} 

While we can 
conclude that smaller mass ratio means a
wider precessional angle, the dependence of the regularity 
of the motion on $\beta $ within that band is not yet clear.
Within the bulk precession, the orbital plane tilts back and forth,
sometimes regularly and sometimes erratically.
Orbits can
subtend roughly the same cone but some occupy the band
more regularly than others.
A hint of the effect of $\beta$ on regularity comes from a comparison 
of frequencies.
An instantaneous orbital frequency can be defined roughly as
	\be
	\omega \sim { { L_N}\over \mu r^2}
	\ee
for comparison with the instantaneous spin precession frequency of the
larger object (neglecting spin-spin coupling just for the crude
estimate)
	\be
	\Omega_1 \sim\left (2+{3m_2\over 2m_1}\right )
	{L_N \over r^3}
	\ee
so that
	\be
	{\omega\over \Omega_1}\simeq 
	{\left (\beta +1 \right )^2\over \beta
	\left (2+{3\beta / 2}\right )}\left ({r\over m}\right ).
	\label{rato}
	\ee
Therefore $\omega/\Omega_1$ is always large.  The pair executes
many orbital windings per spin precession.
This hints that the chaotic behaviour may be related to 
modulation diffusion where a slowly varying parameter, in this case
${\bf \vec S_1}$, facilitates chaotic resonances between coordinates,
in this case $\theta$ and $\phi$ \cite{ll}. 

However, compared to the instantaneous precession frequency of the 
spin of the lighter object
	\be
	{\omega\over \Omega_2}\simeq 
	{\left (\beta +1 \right )^2\over 
	\left (2\beta+{3 / 2}\right )}\left ({r\over m}\right ).
	\label{rato2}
	\ee
This ratio is not as large for small $\beta$ so the lighter companion
always
precesses much more than the heavy black hole.
As argued in \S \ref{second}, the spin of the companion encourages
irregularity and the fact that $\Omega_2/\Omega_1 < 1$ may account for
some of this effect.

In general, the mass ratio $\beta=m_2/m_1$ seems to effect the bulk
shape of the precession and gravitational wave modulation.
By eqns.\ (\ref{rato})
and (\ref{rato2}), $\beta$ may determine the radii at which chaotic
resonances will occur.  
Still, it is unclear how much the mass ratio impacts the
details of the motion.

For comparison with later cases,
the waveform emitted by the BH/BH binary of fig.\ \ref{m.3r20}
is shown
in the top panel of 
fig.\ \ref{spinm.3r20}.  
Even though the precession is fairly regular, it does
modulate the wave amplitude and frequency.  The modulation of 
the waveform has been minimized by placing the 
binary directly above the detector.
Tail-effects are neglected throughout.
A three dimensional view of the precession of the
spin ${\bf \vec S_1}$
is shown in the lower panel of 
fig.\ \ref{spinm.3r20}.

\section{Second Spin}
\label{second}

\begin{figure} 
\centerline{\psfig{file=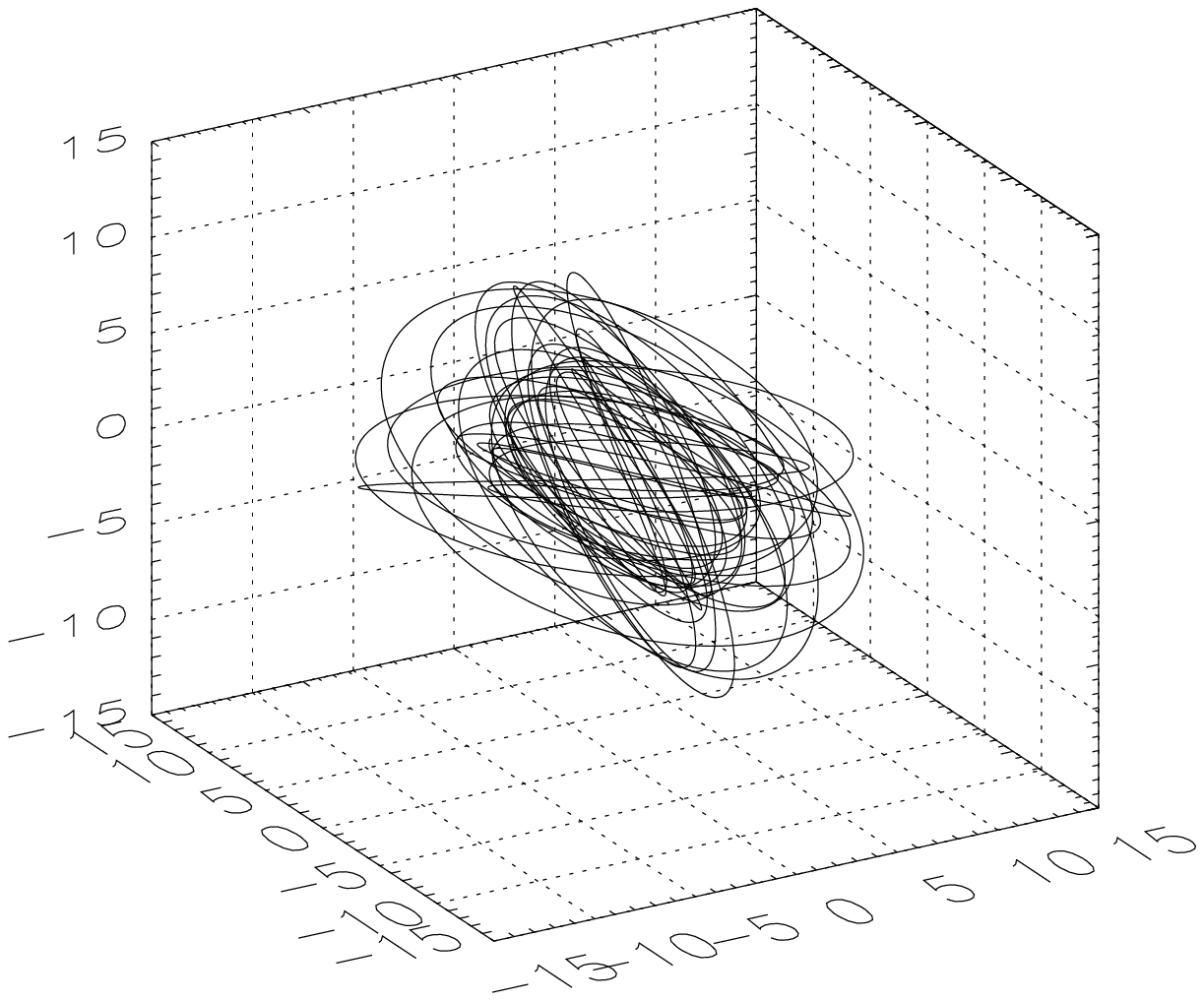,width=3.75in}}
\centerline{\psfig{file=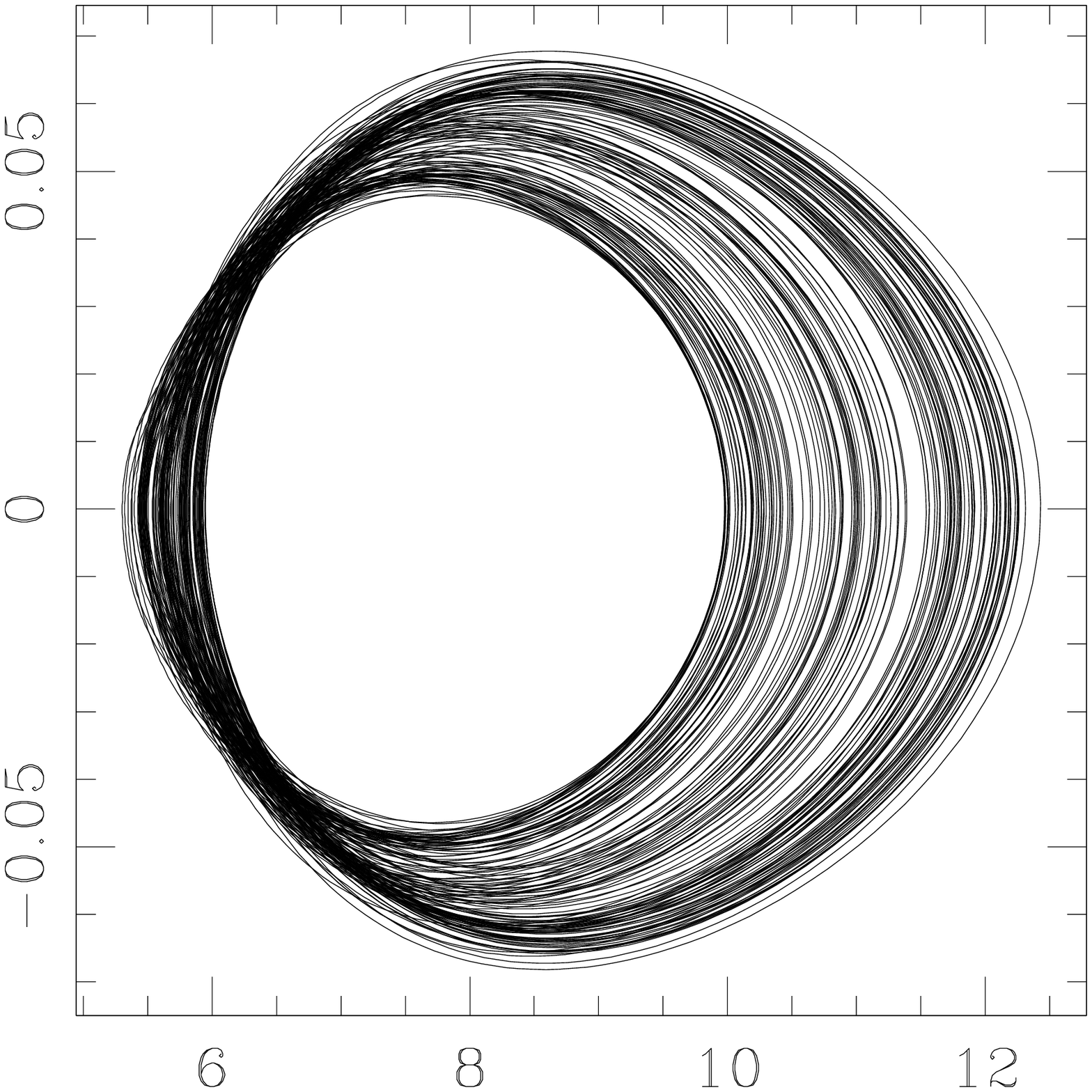,width=2.1in}}
\centerline{\psfig{file=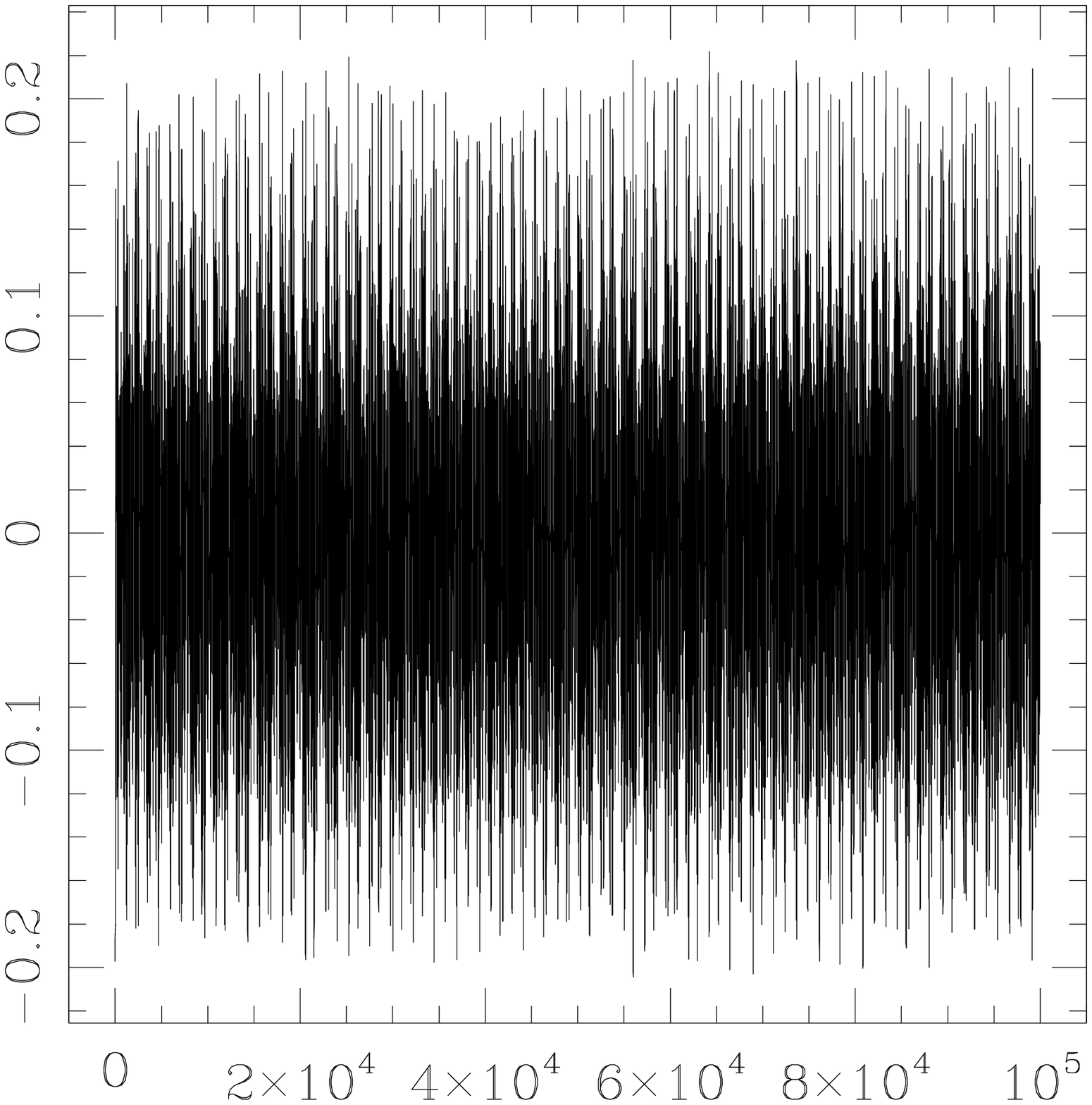,width=2.1in}}
\caption{An irregular orbit with $\beta=m_2/m_1=1/3$, $S_i=m_i^2$.
The initial conditions for the orbit are $r/m=6$ and
$\dot r=0.025$ $r\dot \phi=0.365$. 
Top:  three-dimensional view of the orbit.
Middle:  a projection of the orbit onto the $(r,\dot r)$ plane.
Bottom:  the waveform.
\label{m.3r6s2}}  \end{figure}

\begin{figure}
\centerline{\psfig{file=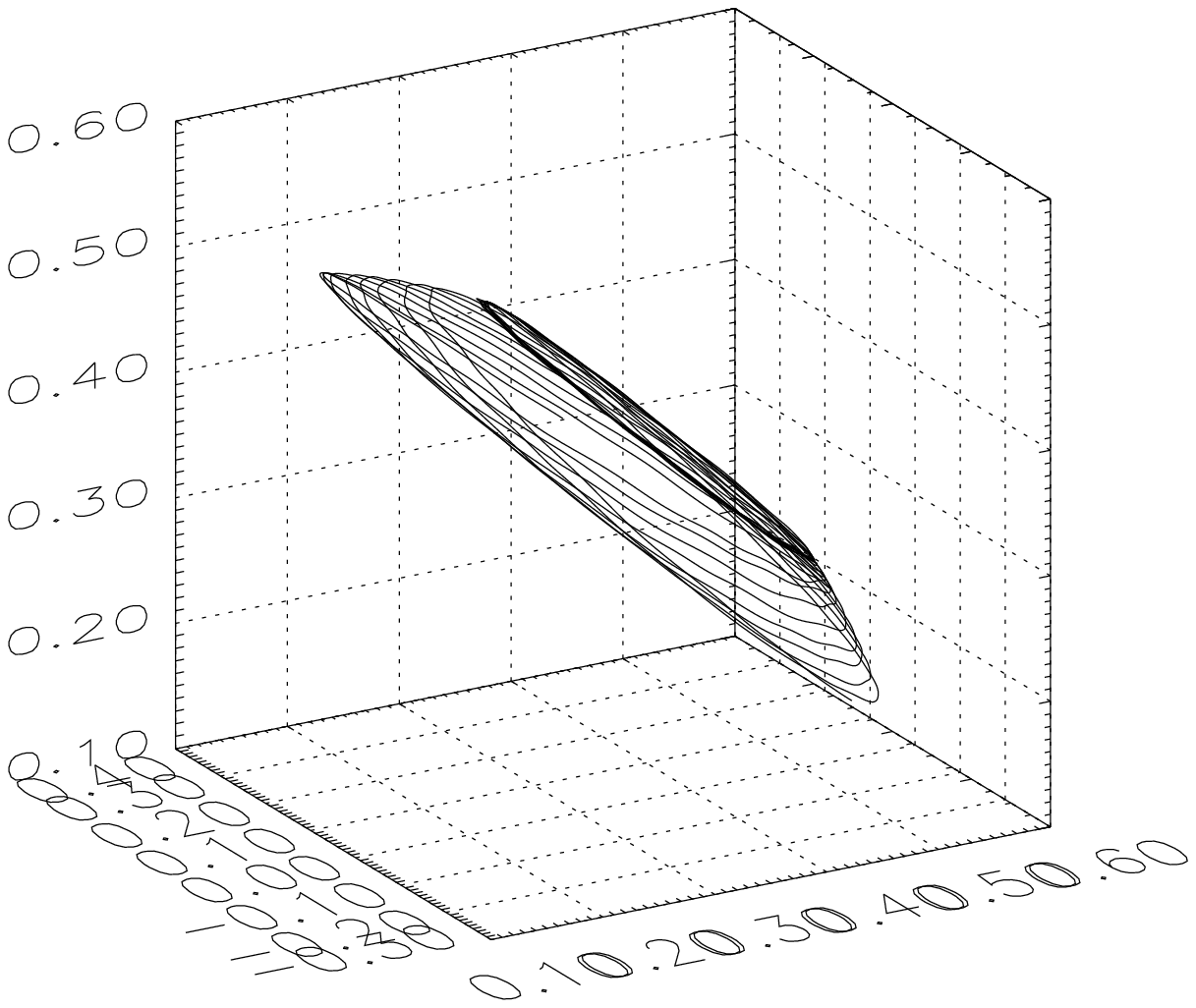,width=3.2in}}
\centerline{\psfig{file=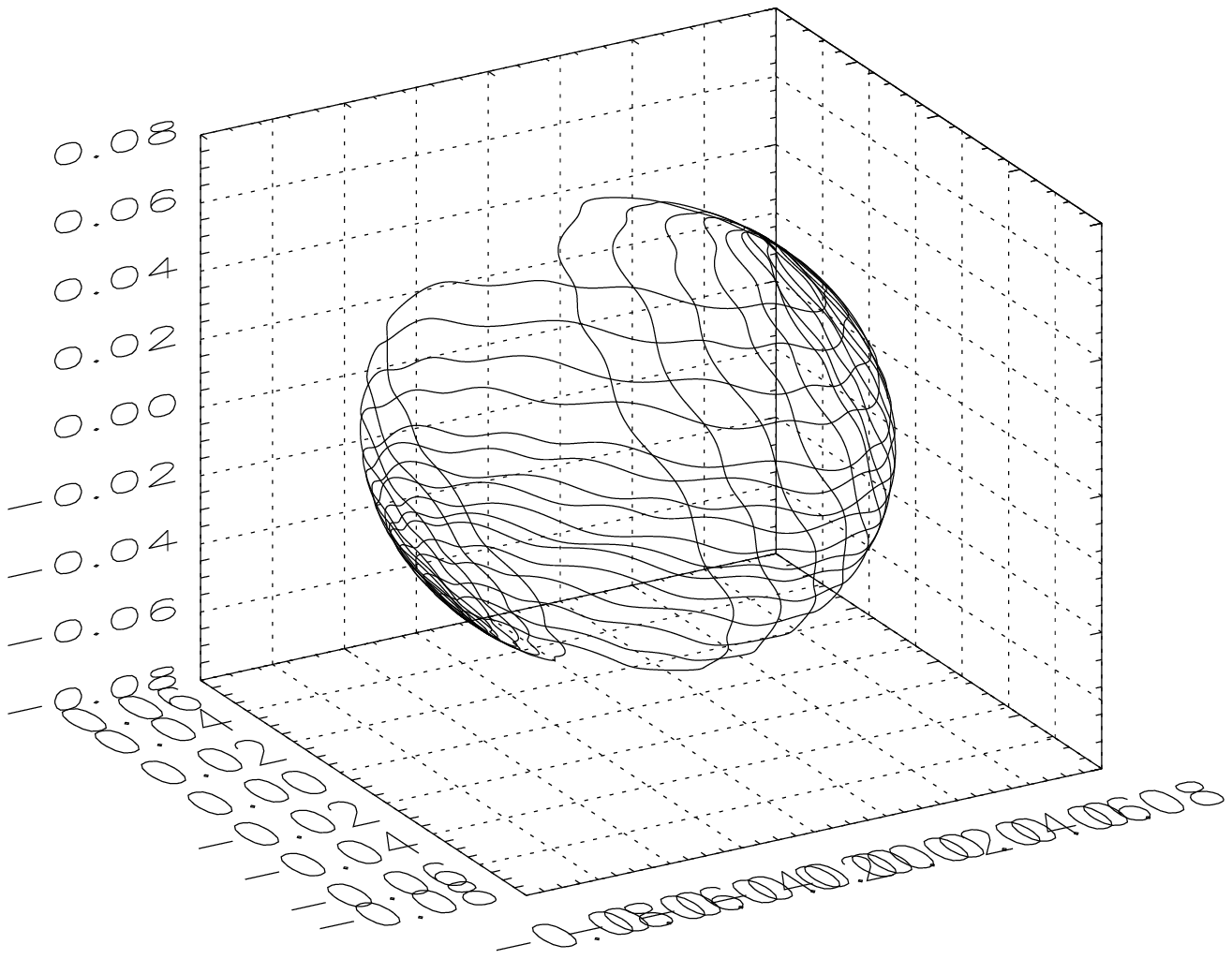,width=2.75in}}
\caption{The Spins
for the orbit of fig.\
\ref{m.3r6s2}.
Top: three-dimensional view of the precession of ${\bf \vec S_1}$.
Bottom:  three-dimensional view of the precession of ${\bf \vec S_2}$.
\label{spinm.3r6s2}}  \end{figure}

The effect of spinning up the lighter companion can be studied by 
starting with the orbits of fig.\ \ref{fbbv}.
Even a small second spin 
will cause further diffusion in phase space.  
When the second spin is maximal, the fraying of the projection 
in $(r,\dot r)$ increases.  Fig.\ \ref{m.3r6s2} shows an irregular 
three-dimensional 
orbit, the projected motion in $(r,\dot r)$, and the waveform
when both objects spin maximally.
Fig.\ \ref{spinm.3r6s2} 
shows the precession of both spin vectors.

The fraying of the orbit in $(r,\dot r)$ indicates that the motion
might not be confined to a torus. As discussed in \S \ref{chaos},
the projection is a hint of nointegrability;
that is, of chaos.  However, as already mentioned, the projection 
alone is 
not enough to conclude there is chaos. A full basin boundary analysis shows
that this orbit occurs near a fractal boundary giving every indication the
orbit is chaotic.  Notice in fig.\ \ref{spinm.3r6s2} that the spin of the
lighter star precesses more than the spin of the heavier object as
expected from eqns.\ (\ref{rato})-(\ref{rato2}).  

\section{Eccentricity}

Binary black holes formed in dense stellar
regions are thought to have
a roughly thermal distribution in eccentricity with 
a slight enhancement of high $e$ at the time of formation
\cite{pzm}.  While many of these may still have time to circularize
before merger, it is worth investigating the role of eccentricity on 
the regularity of the orbit.
Large misaligned spins necessarily induce eccentricity. 
Chaos seems to occur when angular momentum sloshes between spins
and the orbit.
It is difficult to separate cause and effect. Still, it is obvious 
that eccentricity alone is not the culprit.

\begin{figure}
\centerline{\psfig{file=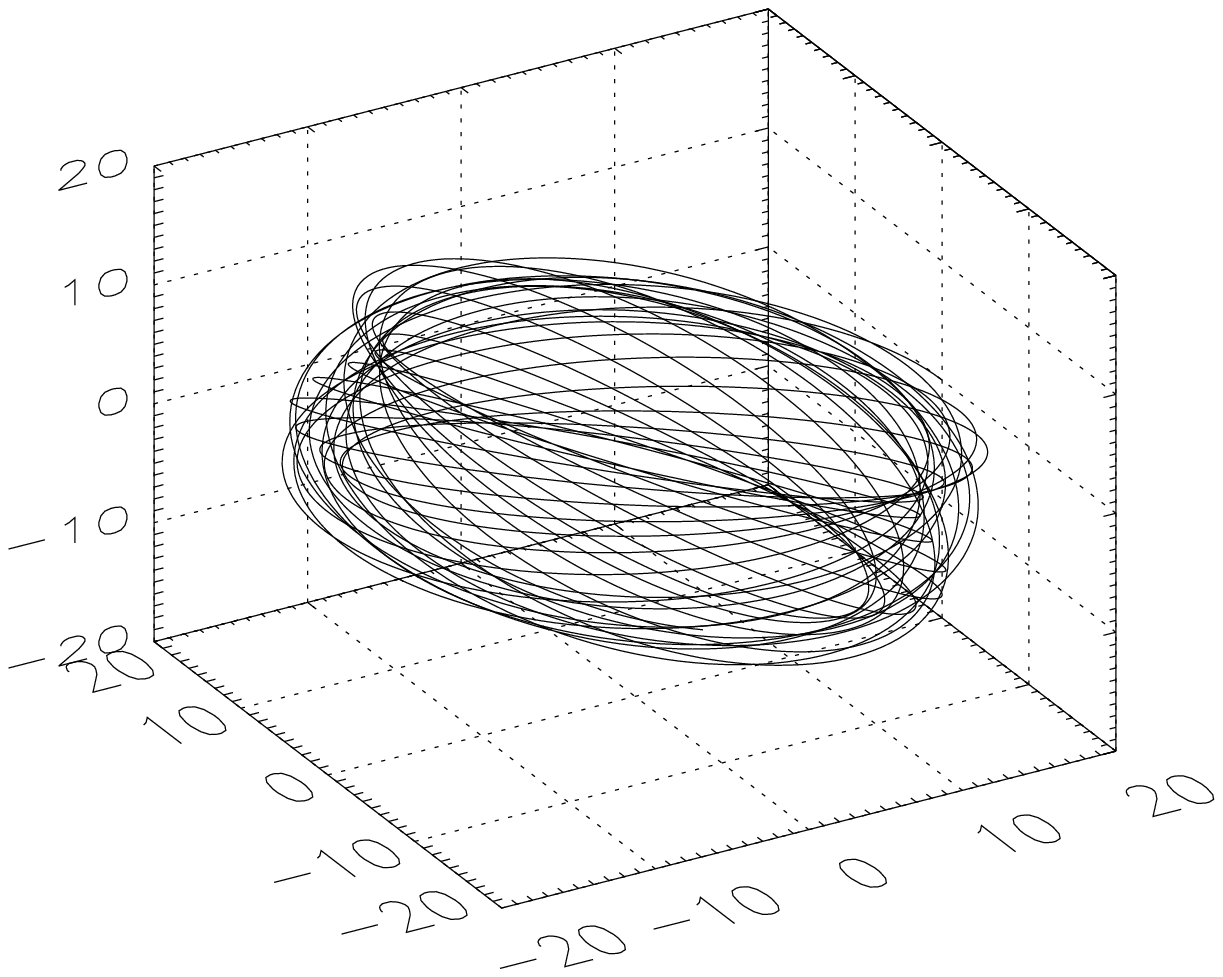,width=4.in}}
\centerline{\psfig{file=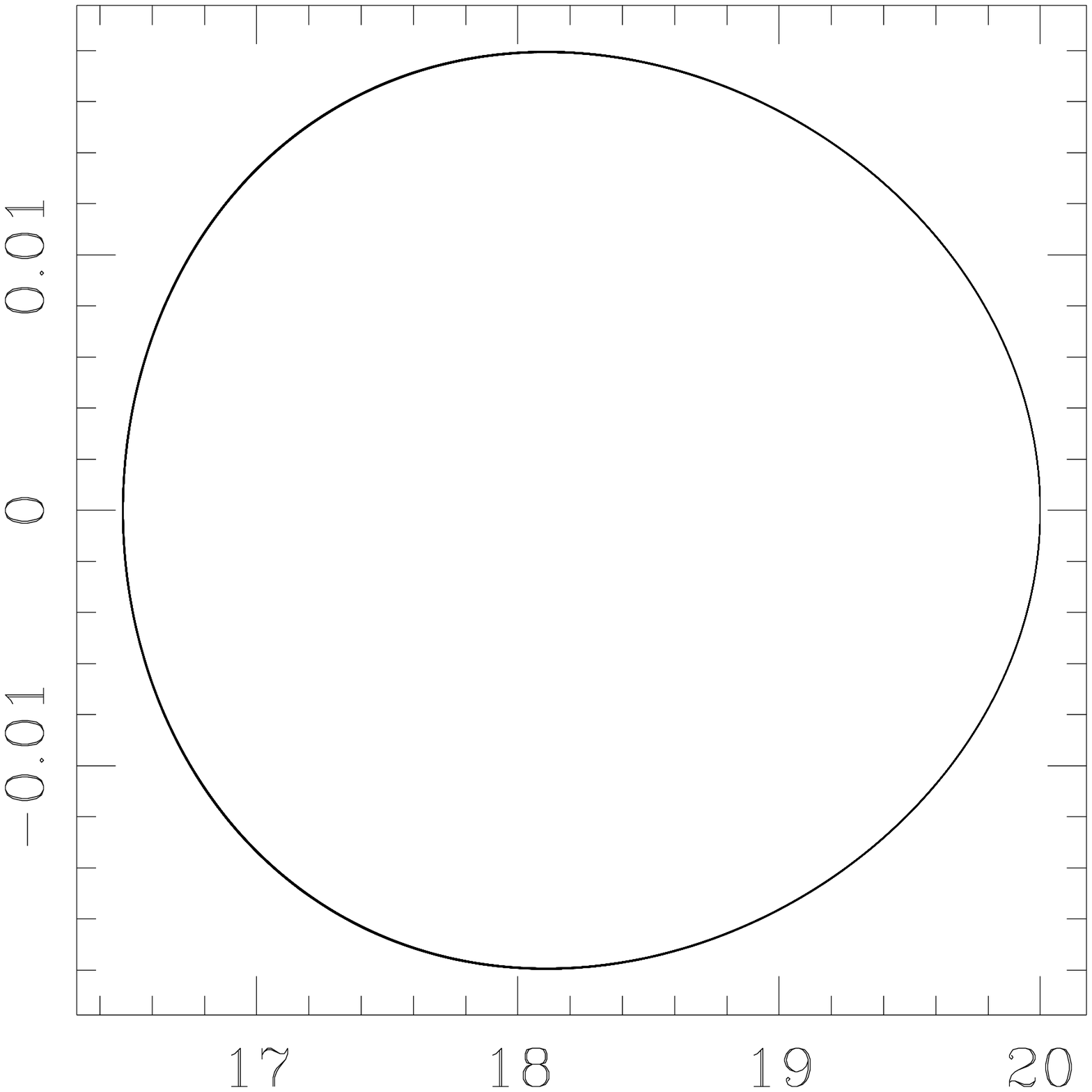,width=2.1in}}
\centerline{\psfig{file=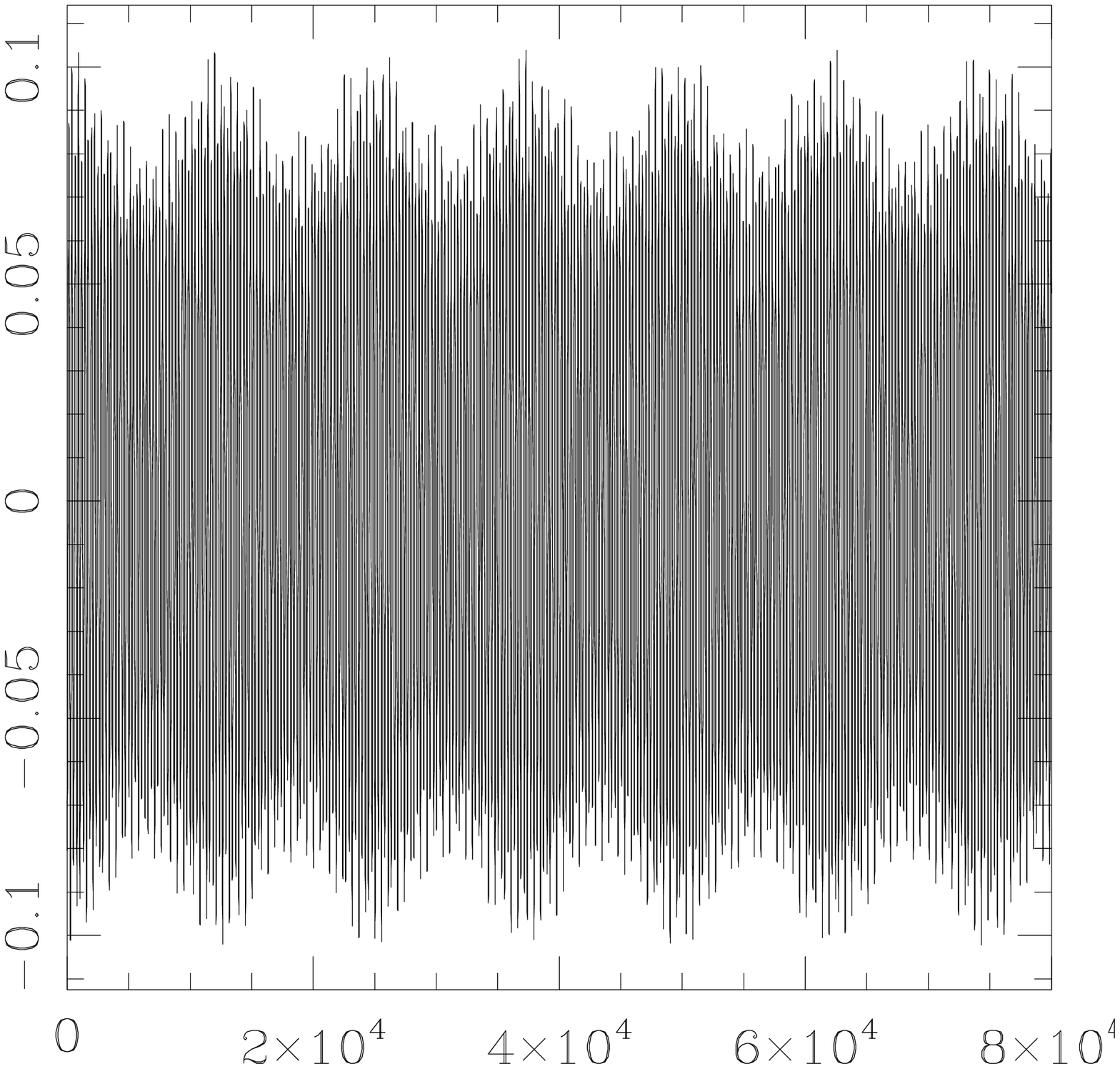,width=2.25in}}
\caption{A regular but eccentric 
orbit with $\beta=m_2/m_1=1/3$, $S_1=m_1^2$ and $S_2=0$.
The initial angle $\theta_1={\rm arccos}({\bf \hat L_N}\cdot{\bf \hat S_1})
= 45{}^o$.  The initial conditions for the orbit are $r/m=20$ and
$r\dot \phi=0.2$.
Top:  three-dimensional view of the orbit.
Middle:  a projection of the orbit onto the $(r,\dot r)$ plane.
Bottom:  the waveform.
\label{m.3r20s1s0_e}}  \end{figure} 

Consider 
fig.\ \ref{m.3r20s1s0_e} where only one of the holes
is spinning.  The top panel shows the three-dimensional
orbit which does have irregular features. However, this is 
deceptive.  It is well known that Keplerian orbits are closed ellipses
while relativistic, elliptical orbits precess within the orbital plane.
The entire plane then precesses due to the spins.
What is being witnessed in the top panel of fig.\ \ref{m.3r20s1s0_e}
is this double precession.  The middle panel shows complete regularity
of the orbit in $(r,\dot r)$.  Motion in this coordinate is confined to 
a torus.  The waveform shown in the bottom panel shows several expected 
features.  Since the orbit is elliptical, the gravitational waves oscillate at
once, twice and three times the orbital frequency which changes the spectrum
from that for a circular orbit.  The double precession then modulates the
amplitude and phase on top of these oscillations.
Even though this motion is regular (for $S_2=0$), in the sense of being
predictable,
the modulation of the waveform from the double precession must
certainly impact observations gained through the method of matched
filtering.  

Eccentric orbits do show chaotic precession when the
companion
spins rapidly as well.  (Note that when both stars spin, there are 
no circular orbits \cite{kidder}.)

\section{Dissipation}
\label{diss}

The emission of gravitational waves dissipates energy and angular momentum.
The spins are not strongly effected by the loss of gravitational waves
which carry away orbital angular momentum and circularize the orbit.
The irregularity of a dissipative binary can be evaluated in terms
of how many orbits the pair executes as it passes through the
successive regions in the conservative system \cite{njco}.  
The most efficient way to do this is again using the fractal basin
boundaries. As pointed out in Ref.\ \cite{njco}, when dissipation is
included the boundaries will never be truly fractal. Like a snowflake,
the self-similar structure will cutoff at some physical limit.
However, if a color coded basin boundary shows substantial structure, 
it is fair to say, the participating orbits are
influenced by irregularity. The smaller the scale at which the cutoff
occurs, the more irregular the history of that set of orbits.

\begin{figure}
\centerline{\psfig{file=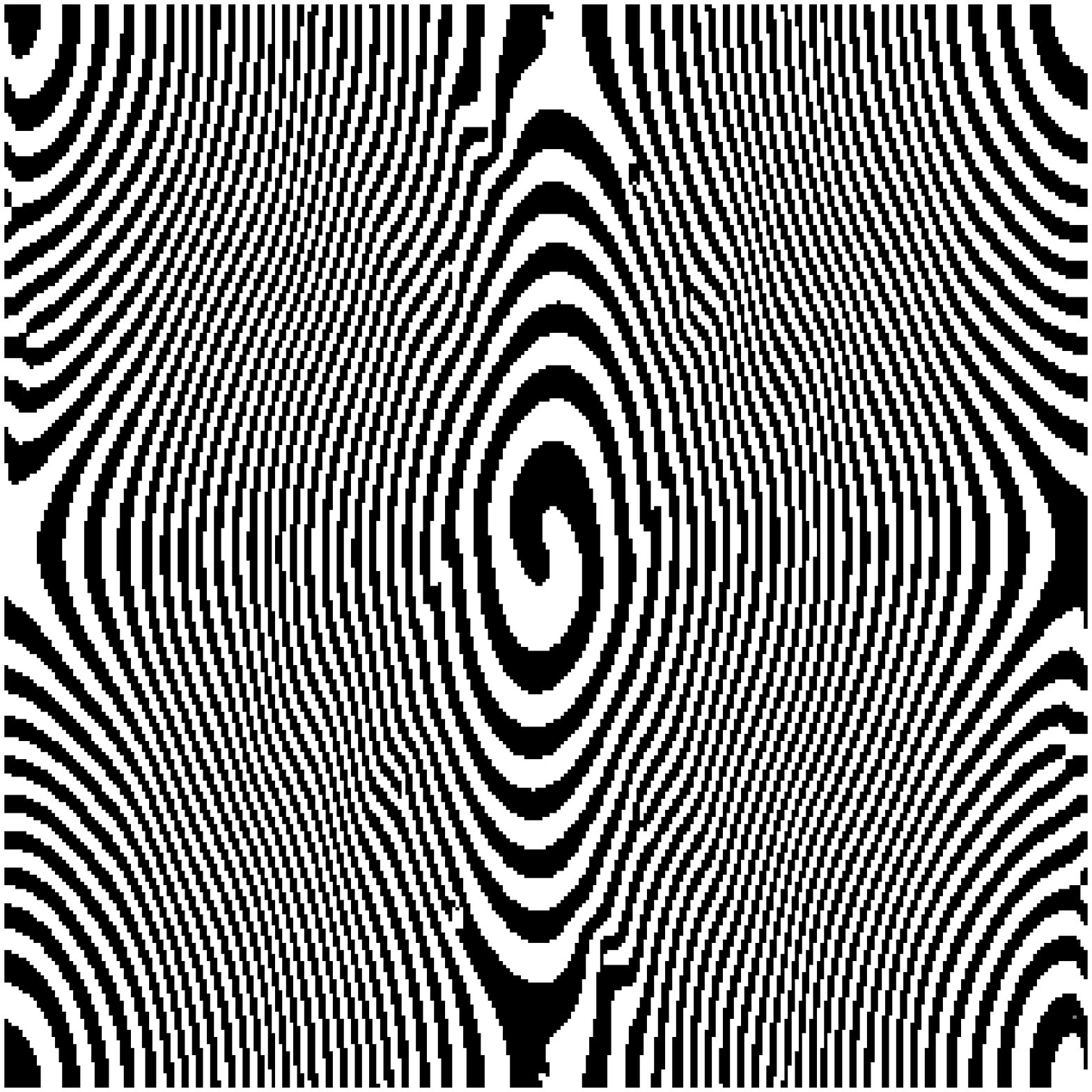,width=2.5in}}
\vskip 15truept
\centerline{\psfig{file=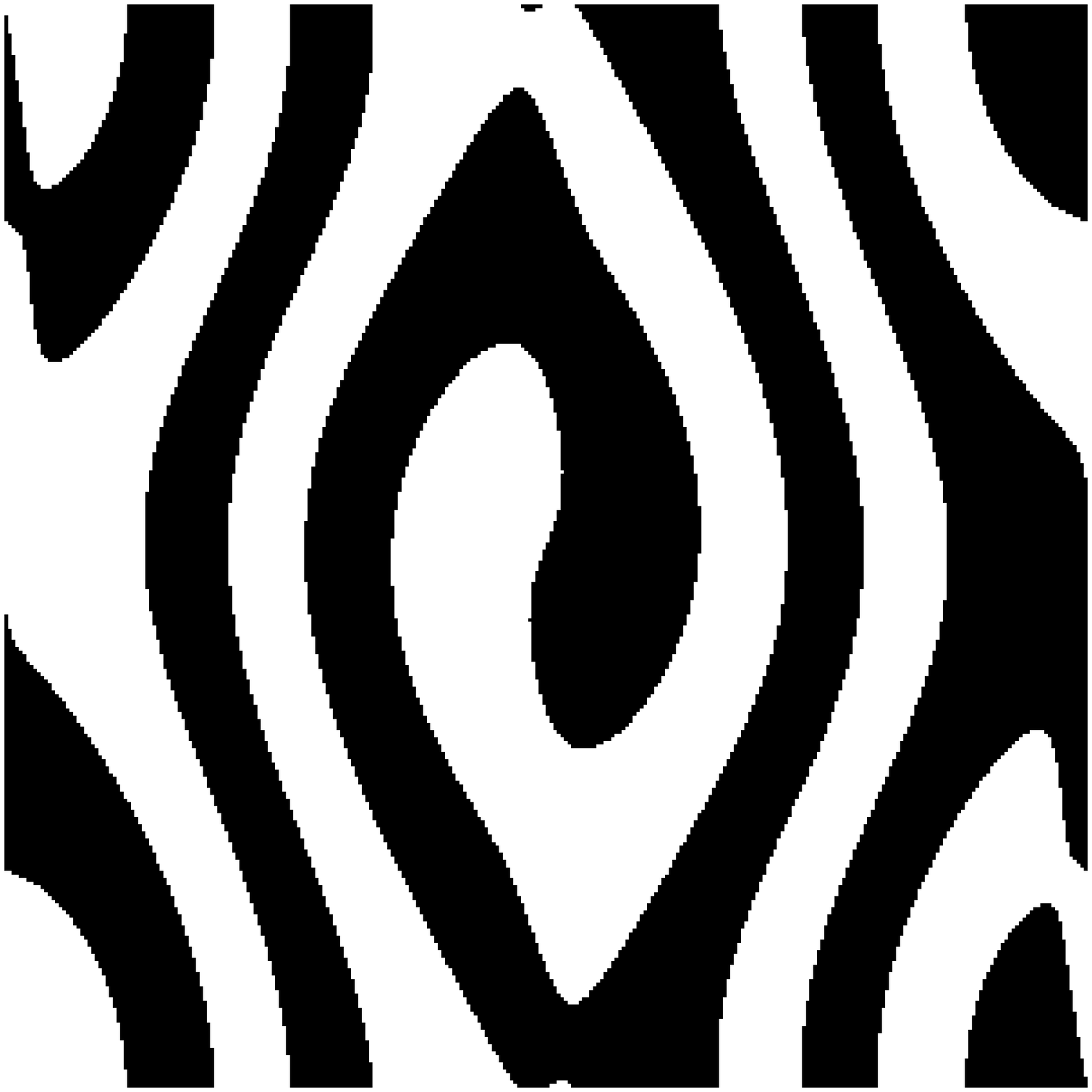,width=2.5in}}
\vskip 10truept
\caption{Coalescing orbits with radiative reaction for
$\beta=1/3$ with $r/m=25$. The slice through phase spans scans
over initial orientation of the spins:
$-180{}^o \le \theta_1 \le 180{}^o$ and $-180{}^o \le \theta_2 \le 180{}^o$.
Upper Panel: Spins are maximal $S_i=m_i^2$.
Lower Panel: Spins are 10 times smaller, $S_i=0.1\ m_i^2$.
$300\times 300$ orbits shown in each panel.
\label{fbbd}}  \end{figure}

Fig.\ \ref{fbbd}  illustrates the 
dissipative inspiral of 90,000 maximally
spinning black hole pairs with $\beta=1/3$.  
The orbits all coalesce due to energy lost to gravitational waves.
The initial condition in $(\theta_1,\theta_2)$
is color coded black if the pair merges from above
the $z$-axis and white if the pair merges from below the
$z$-axis \cite{me1}.
The dissipative system does not provide any ideal outcomes, but this
criterion should not falsely introduce structure.
The top panel of fig.\ \ref{fbbd} shows some structure for maximal
spins while the lower panel of fig.\ \ref{fbbd}
shows substantially less structure for 
spins $1/10$ of maximal.
Incidentally, even more structure is seen for BH/NS binaries
with $\beta=1.4/10$ \cite{me1}.

The orbits begin fairly regularly
although the precession modulates the waveform.
Given unlimited access to theoretical templates,
matched filtering could in principle glean a confident signal, at least
until the pair drew near the unstable orbits.
The orbit becomes more irregular as the separation closes
and the pair passes near 
the  
underlying conservative trajectory
in fig.\ \ref{m.3r6s2}
just before plunge.
Since matched filtering relies on 
the template remaining in phase with the data for many cycles, one
might hope that a disruption of the last few cycles would not be serious.
Many authors have already argued for the use of other methods at
these close separations (see \cite{scott} and the references therein).
However,  others have argued \cite{sathya} that these last few
orbits are heavily weighted and therefore critical for a successful
detection.  More important than the visually
obvious amplitude modulation
the precession modulates
the gravitational wave frequency.  A frequency space analysis is still
required to determine if the modulation inhibits detectability of 
such irregular waves.

\section{Summary}

It is reasonable to conclude for a typical stellar mass
BH/BH system that if the 
primary black hole
spins rapidly, the orbital plane will precess unpredictably for the
final orbits before merger. 
The dependence of the precessional motion was tested as a function 
of three parameters:

$\bullet $ $\beta=m_2/m_1$:  As was already known, the smaller the 
mass ratio, the thicker the band occupied by the precessing orbit
\cite{{acst},{kidder}}.
A ratio of $\beta=1/3$ is small enough for precession to modulate observable
gravitational waves.  The mass ratio may also determine the radius at which 
chaotic resonances occur by effecting the relative orbital and spin
frequencies.

$\bullet $ spin magnitudes:  The transition to chaos occurs as 
the spin magnitudes and misalignments are increased.  There can be 
chaotic motion even if only one body spins.  If it is a light companion
spinning, then the magnitude of the spin has to be larger than maximal
\cite{maeda}.  If it is the heavier object, then chaos can 
occur for rapid but physically allowed black hole spins.
This is true for both BH/BH and BH/NS pairs.
For a given eccentricity and radial separation,
a transition to chaos
can occur as the spin of the companion is increased. 

$\bullet $ eccentricity:  Large misaligned spins cause eccentricity which
may therefore be a feature common to chaotic orbits.
However, large eccentricity alone certainly does not cause chaos.
If chaos is indeed occurring near the unstable homoclinic orbits of
Ref.\ \cite{loc}, then we might expect a fair spread in eccentricities for
chaotic orbits since the homoclinic orbits occur with a broad range
of eccentricities.
Importantly, even for a completely regular orbit
the usual relativistic precession of
an elliptic orbit in the plane becomes superposed on the precession of
the orbital plane.  This combination, though not an indication of chaos,
does lead to a complicated modulation of the gravitational wave signal.

The most pressing question remains, will irregular orbits of coalescing
binaries hinder observations?
A detailed study of the modulated gravitational wave is needed.  A conjecture is
that an erratic precession of the orbital plane will result in an erratic
frequency or phasing of the gravitational wave but will not alter the total
number of cycles much.  If this conjecture is fair, 
observations in the chaotic regime may be salvaged with
a modification of the matched filtering method.
As direct detections become more
acute, we can aspire to watch the transition from regular to irregular
motion.  We might then witness the signature nonlinearity
of general relativity determine the fate of chaotic binaries.

\vskip 25truept

I am grateful to V. Kalogera, V. Kaspi,
S. Portegies Zwart and F. Rasio for sharing their expertise.
I want to especially thank J.D. Barrow, M. Bucher, E. Bertschinger, 
E. Copeland,
N.J. Cornish, P. Ferreira, L. Grischuk, S.A. Hughes, R. O'Reilly,
and B.S. Sathyaprakash
for their insightful comments and support.  I also want to thank
the theoretical physics group at Imperial College for their hospitality.
This work is supported by PPARC.


{}

\appendix

{\hsize\textwidth\columnwidth\hsize\csname
           @twocolumnfalse\endcsname
\widetext
\section{2PN Equations of Motion}

In the notation of Ref. \cite{kidder}, the center of mass equations of
motion in harmonic coordinates are
	\be
	\ddot{ {\bf \vec r} }={\bf \vec a_{PN}} +{\bf \vec a_{SO}}+{\bf \vec
	a_{SS}}+{\bf \vec a_{RR}} \label{eom}
	\ee
with ${\bf \vec r}=r{\bf \hat r}$.
The right hand side is the sum of the contributions to the relative
acceleration from the Post-Newtonian (PN) 
expansion, the spin-orbit (SO)
and
spin-spin (SS) coupling and from the 
radiative reaction (RR).
The explicit terms are quoted from Ref. \cite{kidder}.
The following notation is used:
${{\bf \vec v}}={d{{\bf \vec r}}/dt}$,
${ {\bf \hat n}}\equiv{{{\bf \vec r}}/r}$, $\mu \equiv m_1m_2/m$, $\eta \equiv
\mu /m$, $\delta m \equiv m_1 - m_2$, ${{\bf \vec S}} \equiv {{\bf \vec  S_1}}+{{\bf \vec  S_2}}$,
and ${{\bf \vec  \Delta}} \equiv m({{\bf \vec  S_2}}/m_2 -{{\bf \vec  S_1}}/m_1)$.
The ${\bf \vec a_{PN}}={\bf \vec a_{N}}+{\bf \vec a_{1PN}}+{\bf \vec a_{2PN}}$
with
	\begin{equation}
	{{\bf \vec a_N} }= - {m \over r^2} { {\bf \hat n}} ,
	\end{equation}
	\begin{equation}
	{{\bf \vec a_{1PN}}}=  - {m \over r^2} \left\{  { {\bf \hat n}} \left[
	(1+3\eta)v^2 - 2(2+\eta){m \over r} - {3 \over 2} \eta \dot r^2 \right]
	-2(2-\eta) \dot r {{\bf \vec v}} \right\} ,
	\end{equation}
	\begin{eqnarray}
	{{\bf \vec a_{2PN}}} = - {m \over r^2} \biggl\{ && {{\bf \hat n}} \biggl[ {3 \over 4}
	(12+29\eta) \left ( {m \over r} \right )^2
	+ \eta(3-4\eta)v^4 + {15 \over 8} \eta(1-3\eta)\dot r^4
	\nonumber \\ && \mbox{}
	- {3 \over 2} \eta(3-4\eta)v^2 \dot r^2
	- {1 \over 2} \eta(13-4\eta) {m \over r} v^2
	- (2+25\eta+2\eta^2) {m \over r} \dot r^2 \biggr]
	\nonumber \\ && \mbox{}
	- {1 \over 2} \dot r {{\bf \vec v}}
	\left[ \eta(15+4\eta)v^2 - (4+41\eta+8\eta^2)
	{m \over r} -3\eta(3+2\eta) \dot r^2 \right] \biggr\} ,
	\end{eqnarray}
The radiative reaction term is due to terms to 
$5/2$PN order and can be expressed as
	\begin{equation}
	{\bf \vec a_{RR}} = {8 \over 5} \eta {m^2 \over r^3} \left\{ \dot r { {\bf \hat n}}
	\left[ 18v^2 + {2 \over 3} {m \over r} -25 \dot r^2 \right] - {{\bf \vec v}}
	\left[ 6v^2 - 2
	{m \over r} -15 \dot r^2 \right] \right\} ,
	\end{equation}
The spin-orbit acceleration is
	\begin{equation}
	{\bf \vec a_{SO}} = {1 \over r^3} \left\{ 6 {{\bf \hat n}} [( { {\bf \hat n}} \times
	{{\bf \vec v}} ) { \cdot} (2{{\bf \vec S} }+ {\delta m \over m}{{\bf \vec \Delta}} )]
	- [ {{\bf \vec v}} \times (7
	{{\bf \vec S}}+3{\delta m \over m}{{\bf \vec \Delta}})]
	+ 3 \dot r [ { {\bf \hat n}} \times
	(3{{\bf \vec S}} + {\delta m \over m}{{\bf \vec \Delta}} )] \right\} .
	\end{equation}
and the spin-spin acceleration is
	\begin{equation}
	{\bf \vec a_{SS}} = - {3 \over \mu r^4} \biggl\{ {{\bf \hat n}} ({{\bf \vec S_1 }\cdot
	{\bf \vec S_2}}) + {{\bf \vec S_1}} ({{\bf \hat n} \cdot {\bf \vec S_2}}) + {{\bf \vec S_2}} 
	({{\bf \hat n }\cdot
	{\bf \vec S_1}}) - 5 {{\bf \hat n}} ({ {\bf \hat n} \cdot {\bf \vec S_1}})({{\bf \hat n}
	\cdot {\bf \vec
	S_2}})
	\biggr\} .
	\end{equation}

\subsection{Constants of the Motion}

	\begin{equation}
E =  E_{PN} + E_{SO} + E_{SS} ,
\end{equation}
where
$E_{PN}=E_N+E_{1PN}+E_{2PN}$ and
	\begin{equation}
	E_N = \mu \left\{ {1 \over 2} v^2 - {m \over r} \right\} ,
	\end{equation}
	\begin{equation}
	E_{1PN} =  \mu \left\{
	{3 \over 8}
	(1-3\eta) v^4
	+{1 \over 2} (3+\eta) v^2 {m \over r}  +
	{1 \over 2} \eta {m \over r} \dot r^2 + {1 \over 2} ({m \over r})^2 \right\} ,
	\end{equation}
	\begin{eqnarray}
	E_{2PN} = \mu \biggl\{ && {5 \over 16}(1-7\eta+13\eta^2) v^6
	- {3 \over 8} \eta (1-3\eta){m \over r} \dot r^4
	+ {1 \over 8} (21-23\eta-27\eta^2) {m \over r} v^4 \nonumber \\ && \mbox{}
	+ {1 \over 8} (14-55\eta+4\eta^2) \left( {m \over r} \right)^2 v^2
	+ {1 \over 4} \eta (1-15\eta) {m \over r}
	v^2 \dot r^2
	 - {1 \over 4} (2+15\eta) \left( {m \over r} \right) ^3 \nonumber \\ && \mbox{}
	+ {1 \over 8}
	(4+69\eta+12\eta^2) \left( {m \over r} \right) ^2 \dot r^2 \biggr\} ,
	\end{eqnarray}
	\begin{equation}
	E_{SO} = {1 \over r^3} {{\bf \vec  L_N }\cdot} ({{\bf \vec  S}} + {\delta m \over m}
	{{\bf \vec  \Delta}}) ,
	\end{equation}
	\begin{equation}
	E_{SS} = {1 \over r^3} \left\{ 3 \left( { {\bf \hat n} \cdot {\bf \vec S_1}} \right)
	  \left( { {\bf \hat n} \cdot{\bf \vec S_2}} \right) - \left( {{\bf \vec  S_1}
\cdot {\bf \vec S_2}}
	  \right) \right\} .
	\end{equation}
The total angular momentum is given by
	\begin{equation}
	{{\bf \vec  J}} = {{\bf \vec  L} }+ {{\bf \vec  S}} ,
	\end{equation}
	where
	
	\begin{equation}
	{{\bf \vec  L}} = {\bf \vec  L_{PN}} + {\bf \vec  L_{SO}}
	\label{split}
	\end{equation}
with 
${\bf \vec L_{PN}}={\bf \vec L_{N}}+{\bf \vec L_{1PN}}+{\bf \vec L_{2PN}}$ 
and
		\ba
	{{\bf \vec  L_N}} \equiv \mu ({{\bf \vec  r} \times {\bf \vec v}})
	,
	\ea
	\begin{equation}
	{\bf \vec  L_{1PN}} = {{\bf \vec  L_N}} \left\{
	{1 \over 2} v^2
	(1-3\eta) + (3+\eta) {m \over r} \right\} ,
	\label{split1}
	\end{equation}
	\begin{eqnarray}
	{\bf \vec  L_{2PN} }= {{\bf \vec  L_N} }\biggl\{ &&
	 {3 \over 8} (1-7\eta+13\eta^2) v^4
	- {1 \over 2}\eta (2+5\eta) {m \over r} \dot r^2 \nonumber \\ && \mbox{}
	+ {1 \over 2}
	(7-10\eta-9\eta^2) {m \over r}v^2 \nonumber \\ && \mbox{}
	+ {1 \over 4} (14-41\eta+4\eta^2)
	\left( {m \over r} \right)^2 \biggr\} ,
		\label{split2}
	\end{eqnarray}
and
	\begin{eqnarray}
	{\bf \vec  L_{SO}} = {\mu \over m} \Biggl\{ &&
	{m \over r} {{\bf \hat n} \times} \left[ {{\bf \hat n} \times} \left(
	3 {{\bf \vec  S}} + {\delta m \over m}{{\bf \vec  \Delta}} \right) \right] \nonumber \\
	&& \mbox{} - {1 \over 2}
	{{\bf \vec  v }\times} \left[ {{\bf \vec  v} \times} \left( {{\bf \vec  S}} + {\delta m \over m}
	{{\bf \vec  \Delta}} \right) \right] \Biggr\} .
	\label{split3}
	\end{eqnarray}

}

\end{document}